\documentclass[aps,prc,twocolumn,superscriptaddress,nofootinbib]{revtex4-1}

\usepackage{amssymb}
\usepackage{amsmath}
\usepackage{siunitx}
\usepackage{pifont}
\usepackage{multirow}
\usepackage{color}

\usepackage{hyperref}
\hypersetup{letterpaper, colorlinks=true, breaklinks=True, citecolor=blue}
\usepackage{nicefrac}

\usepackage{graphicx}

\makeatletter

\newcommand{\Rmnum}[1]{\expandafter\@slowromancap\romannumeral #1@}
\makeatother

  \newcommand {\nc} {\newcommand}
  \nc {\beq} {\begin{eqnarray}}
  \nc {\eeq} {\nonumber \end{eqnarray}}
  \nc {\eeqn}[1] {\label {#1} \end{eqnarray}}
  \nc {\eol} {\nonumber \\}
  \nc {\eoln}[1] {\label {#1} \\}
  \nc {\ve} [1] {\mbox{\boldmath $#1$}}
  \nc {\ves} [1] {\mbox{\boldmath ${\scriptstyle #1}$}}
  \nc {\mrm} [1] {\mathrm{#1}}
  \nc {\half} {\mbox{$\frac{1}{2}$}}
  \nc {\thal} {\mbox{$\frac{3}{2}$}}
  \nc {\fial} {\mbox{$\frac{5}{2}$}}
  \nc {\la} {\mbox{$\langle$}}
  \nc {\ra} {\mbox{$\rangle$}}
  \nc {\etal} {\emph{et al.}}
  \nc {\eq} [1] {(\ref{#1})}
  \nc {\Eq} [1] {Eq.~(\ref{#1})}
  \nc {\Sec} [1] {Sec.~\ref{#1}}
  \nc {\chap} [1] {Chapter~\ref{#1}}
  \nc {\anx} [1] {Appendix~\ref{#1}}
  \nc {\tbl} [1] {Table~\ref{#1}}
  \nc {\Fig} [1] {Fig.~\ref{#1}}
  \nc {\ex} [1] {$^{#1}$}
  \nc {\Sch} {Schr\"odinger }
  \nc {\flim} [2] {\mathop{\longrightarrow}\limits_{{#1}\rightarrow{#2}}}
  \nc {\IR} [1]{\textcolor{red}{#1}}
  \nc {\IB} [1]{\textcolor{blue}{#1}}
  \nc{\IG}[1]{\textcolor{green}{#1}}

\begin{document}

\title{First experimental test of the ratio method for nuclear-reaction analysis}

\author{S. Ota}
\email{sota@bnl.gov}
\affiliation{National Nuclear Data Center, Brookhaven National Laboratory, Upton, NY 11973-5000, USA}
\affiliation{Cyclotron Institute, Texas A\&M University, College Station, TX 77843, USA}
\author{P. Capel}
\email{pcapel@uni-mainz.de}
\affiliation{Institut f\"ur Kernphysik, Johannes Gutenberg-Universit\"at Mainz, D-55099 Mainz, Germany}
\author{G. Christian}
\affiliation{Department of Astronomy \& Physics, Saint Mary's University, Halifax, NS B3H~3C3, Canada}
\affiliation{Department of Physics \& Astronomy, Texas A\&M University, College Station, TX 77843, USA}
\author{V. Durant}
\affiliation{Institut f\"ur Kernphysik, Johannes Gutenberg-Universit\"at Mainz, D-55099 Mainz, Germany}
\author{K. Hagel}
\affiliation{Cyclotron Institute, Texas A\&M University, College Station, TX 77843, USA}
\author{E. Harris}
\affiliation{Cyclotron Institute, Texas A\&M University, College Station, TX 77843, USA}
\affiliation{Department of Physics \& Astronomy, Texas A\&M University, College Station, TX 77843, USA}
\author{R. C. Johnson}
\affiliation{School of Mathematics and Physics, University of Surrey, Guildford, Surrey, GU2 7XH, United Kingdom}
\author{Z. Luo}
\affiliation{Cyclotron Institute, Texas A\&M University, College Station, TX 77843, USA}
\affiliation{Department of Physics \& Astronomy, Texas A\&M University, College Station, TX 77843, USA}
\author{F. M. Nunes}
\affiliation{Facility for Rare Isotope Beams, Michigan State University, East Lansing, Michigan 48824, USA}
\affiliation{Department of Physics and Astronomy, Michigan State University, East Lansing, Michigan 48824, USA}
\author{M. Roosa}
\affiliation{Cyclotron Institute, Texas A\&M University, College Station, TX 77843, USA}
\affiliation{Department of Physics \& Astronomy, Texas A\&M University, College Station, TX 77843, USA}
\author{A. Saastamoinen}
\affiliation{Cyclotron Institute, Texas A\&M University, College Station, TX 77843, USA}
\author{D. P. Scriven}
\affiliation{Cyclotron Institute, Texas A\&M University, College Station, TX 77843, USA}
\affiliation{Department of Physics \& Astronomy, Texas A\&M University, College Station, TX 77843, USA}



\date{\today}
 
\begin{abstract}
Nuclear halos are exotic quantal structures observed far from stability.
They are mostly studied through reactions.
The ratio of angular cross sections for breakup and scattering is predicted to be independent of the reaction process and to be very sensitive to the halo structure.
We test this new observable experimentally for the first time on the collision of $^{11}$Be on C at 22.8 MeV/nucleon and using existing data on Pb at 19.1 MeV/nucleon.
The theoretical predictions are verified, which offers the possibility to develop a new spectroscopic tool to study nuclear structure far from stability.
\end{abstract}


\maketitle

Collisions are a powerful tool in many fields of quantum physics.
They provide key information about the interactions between particles (molecules, atoms, nuclei, hadrons\ldots).
They are also used to measure the rate of reactions that have a wide range of interests: from fundamental physics, such as astrophysics, to applied physics, such as nuclear engineering.
Finally, collisions are sometimes the only way to study exotic quantal structures.
This is particularly true in the realm of nuclear physics, where the development of radioactive ion beams in the 1980s has enabled the study of nuclei far from stability, leading to the discovery of \emph{halo nuclei} \cite{Tan85l,Tan85b}.

Halo nuclei exhibit an unusually large size compared to their isobars, breaking the empirical rule that nuclear radii scale with $A^{1/3}$, the cubic root of the mass number.
They are located near the neutron dripline.
In this extremely unstable region of the nuclear chart, valence neutrons can be very loosely bound to the nuclei.
Therefore, they can tunnel far into the classically forbidden region to form a diffuse halo surrounding a dense and compact core \cite{HJ87}.
This exotic structure has been the subject of many studies, both experimental and theoretical \cite{Tan96}.

Because of their short lifetime, halo nuclei are mostly studied through reactions.
Their exceptional spatial extension was first evidenced thanks to the large interaction cross sections measured in collisions with various targets \cite{Tan85l,Tan85b,Tan96}.
In knockout reactions, the halo neutrons are removed at high energies on a light target.
The presence of a halo leads to a narrow momentum distribution of the outgoing core \cite{Aum00,Tan96}.
A more exclusive measurement is the diffractive breakup, where both the core and the valence neutrons are measured in coincidence \cite{Pal03,Fuk04}.
The corresponding cross section is large due to the fragile nature of the halo structure.
In elastic scattering, the significant coupling to breakup strongly affects the differential cross section \cite{DiP10,DiP12}.
Halo states can also be directly populated via $(d,p)$ transfer reactions \cite{Sch12,Sch13}.

All these reactions provide valuable information about halo nuclei.
Unfortunately, the detailed analysis of experimental data is hampered by the sensitivity of reaction calculations to the optical potentials, which account for the interaction between the projectile constituents and the target \cite{CGB04,SNP22,HWL23,CLN23}.
The structure information inferred from experiment seems inherently marred by the uncertainty of the input of the reaction models.

A detailed analysis of differential cross sections for elastic scattering and breakup of halo nuclei shows that both processes exhibit very similar diffraction patterns \cite{CHB10}, indicating that the projectile is scattered in a similar way whether it remains bound or if it dissociates.
This is easily explained within the Recoil Excitation and Breakup model (REB) \cite{JAT97,proc97}.
In that model, the differential cross sections for both scattering and breakup factorize as the product of the cross section for a pointlike projectile times slowly varying form factors that account for the extension of the halo.

In addition to explaining the similarity of these diffraction patterns, the REB brings a new idea in the study of reactions involving exotic nuclei.
Taking the ratio of these cross sections should remove the strong dependence on the reaction process and on the projectile-target interaction.
This ratio is also predicted to be function of form factors of the projectile wave functions.
It should therefore depend only on the projectile structure \cite{CJN11,CJN13,CJN20}.
Moreover, corresponding to the ratio of cross sections, it does not require a precise normalization of the data, a precious experimental advantage.
Accordingly, this new reaction observable should give direct access to the nuclear structure of the projectile: its one-neutron separation energy, the asymptotic normalization coefficient (ANC), the core-neutron orbital angular momentum $\ell$ 
and even precise information on the overlap wave function such as its root-mean-square radius \cite{CJN11,CJN13}.

Using accurate models of the collision, various theoretical studies have confirmed the  sensitivity of the ratio to the projectile structure \cite{CJN13,CCN16,YCP19} and demonstrated its applicability beyond the original idea of Ref.~\cite{CJN11}. 
First the method can be used to study neutron halos over a wide range of beam energies starting from at least 20~MeV/nucleon \cite{CCN16}.
Second, it can be applied to proton halos as well \cite{YCP19}.
Finally, it is not limited to halo nuclei, and can be used to infer structure information about valence nucleons with a large separation energy or bound in an orbital with $\ell\ge2$  \cite{CJN13}.
The ratio method therefore provides a new reaction observable to study nuclear structure far from stability.

Until now, the ratio method has not been experimentally verified because the breakup and scattering channels have not yet been measured at the same beam energy on the same target. 
In this letter, we report the first simultaneous measurement of both channels for $^{11}$Be impinging on C at 22.8~MeV/nucleon.
We analyze these new data in the framework of the ratio method.
We extend this analysis to the $^{11}$Be-Pb collision at 19.1~MeV/nucleon recently measured in Lanzhou by Duan \etal\ \cite{Duan22}.

The experiment was performed at Texas A\&M University (TAMU) using the K500 superconducting cyclotron. 
A $^{11}$Be beam was produced by bombarding a 1 mm thick $^{9}$Be target with a 30 MeV/nucleon $^{13}$C primary beam. 
The fragments were selected with a dipole magnet and a velocity filter following the production (momentum acceptance $\delta p/p=\pm 0.66$\% (FWHM)), and then $^{11}$Be (4$^+$) ions were separated from other isotopes using the MARS dipole magnet \cite{Tribble1989}. 
The $^{11}$Be beam with the energy of $22.8 \pm 0.3$ MeV/nucleon was delivered to the target chamber. 
The profile and purity of the $^{11}$Be beam have been measured at a reduced beam rate using a Si $\Delta E$-$E$ telescope with an active area of $5\times5$ cm$^2$ placed approximately 45 cm upstream from the reaction target. 
The $\Delta E$ detector is resistive along the MARS's dispersive axis ($y$) and segmented by 16 strips along the non-dispersive axis ($x$), leading to the position resolution of approximately $\delta y = 0.2$ mm and $\delta x = 3.1$ mm. 
The beam was limited with a slit ($\pm 5$ mm in $y$ and $\pm 25$ mm in $x$) placed $\sim 10$ cm upstream from the telescope to block other beam species by limiting the magnetic rigidity. 
The beam spot size was measured with the Si telescope to be $\sigma_{x}\sim 3$ mm and $\sigma_{y}\sim 2$ mm. 
The beam spot size remained nearly the same at the reaction target position as described below. 
The Si detectors were then removed from the beamline and the total intensity of the beam was determined with a phoswich detector to be typically $1.2 \times10^4$ particles per second (pps), among which the $^{11}$Be beam intensity was about $7.2 \times10^3$ pps. 
Thus, the $^{11}$Be beam purity was determined to be about 60\% with 6\% from $^8$Li and the remaining from not well-focused light ions. 

For the scattering/reaction target, we used natural carbon in the form of flexible graphite with 99.8\% purity with a thickness of 8.7 (3) mg/cm$^2$.
The thickness was determined from a weighted average of the values measured by two different weighing scales, a caliper, and the energy loss of $^{11}$Be scattering events.
To check the scattering yields and the thickness of the target from the energy loss, two other targets with thickness 17.4 (6) mg/cm$^2$ and 34.8 (12) mg/cm$^2$ were also used during the experiment. 
Consistency of the yields after normalization to the incident beam counts was confirmed within the uncertainty of the target thicknesses. 
The target size was limited to a 17 mm diameter and the incident beam outside the area was blocked by an aluminum target holder.

The BlueSTEAl detector system \cite{Ota2023} was used to measure the scattering and breakup cross sections in the $^{11}$Be-C collision. 
Because the removed neutron was not measured, the latter corresponds to inclusive breakup.
This detector comprises four layers of Micron S2 Si DSSD (one 0.5 mm thick and three 1.5 mm thick) and a phoswich plastic scintillator with 1.1 cm thickness. 
Two $\Delta E$-$E$ telescopes were built from the four Si DSSDs [$\rm 0.5~mm+1.5$ mm (Si 1\&2) and two 1.5 mm (Si 3\&4)] (see Fig.~13(d) in Ref.~\cite{Ota2023}). 
In the first part of the experiment, we placed the Si 1\&2 telescope at 5.08 cm and the Si 3\&4 telescope at 17.78 cm downstream from the target [Config.\ 1].
In the second part of the experiment, we placed the Si 3\&4 telescope at 10.16 cm downstream from the target [Config.\ 2] while the Si 1\&2 telescope was placed 2.54 cm upstream from the target to monitor the beam's spatial spread and purity and to shield the Si 3\&4 telescope from the defocused beam. 
As discussed in detail in Ref.~\cite{Ota2023}, the measurements with the two configurations supplement each other, enabling us to double check the consistency of the measured cross sections in the overlapping angular range. 
Since the relatively broad beam-spot size causes poor angular resolution for the Si 1\&2 detectors in Config.\ 1,  it was difficult to extract  precise angular cross sections.
Therefore we have not used the data from that telescope.

 \begin{figure}[t]
        \centering
          \includegraphics[width=8cm]{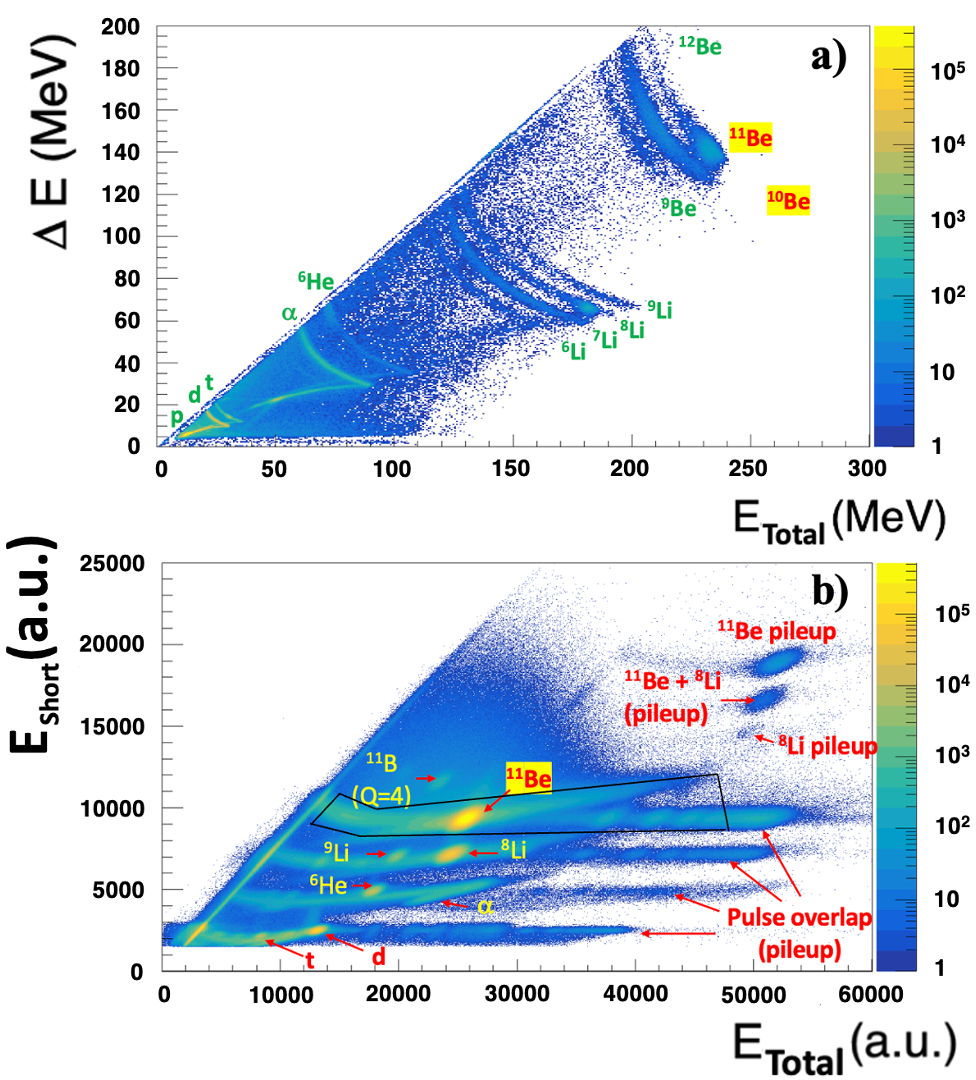}    
        \caption
        {(a) PID by $\Delta E$-$E$ plot from the downstream Si telescope (Config.\ 1; 14 h).
        (b) PID by $E_{\rm short}$ vs $E_{\rm total}$ plot with the phoswich detector.
        }\label{fig:Fig2}
\end{figure}

As shown in Fig.~\ref{fig:Fig2}(a), a clear particle identification (PID) was obtained with the Si 3\&4 telescope.
The numbers of $^{11}$Be and $^{10}$Be detected at different angles were used to obtain differential cross sections for both scattering and inclusive breakup.
The beam offset from the telescope's center was deduced from the scattering measured at different azimuthal angles as explained in Ref.~\cite{Ota2023}.
It was typically a few mm in $x$ and $y$ directions with an uncertainty $\Delta x,y$$<$($0.5$,$0.5$)~mm.
During the experiment, background measurements were performed with a blank target and also without the target holder. 
Negligible ($<1$\%) background $^{11,10}$Be events measured with the blank target 
ensure that the observed $^{11,10}$Be came from scattering/reaction with the target. 
The phoswich detector was placed about 30 cm downstream from the target to count the beam rate with clear elemental identification [Fig.~\ref{fig:Fig2}(b)]. 
The PID in the phoswich's $E_{\rm short}$ vs $E_{\rm total}$ plot was made with the help of some punch-through events from the Si telescope (e.g., $^8$Li, $\alpha$-particles, deuterons and protons; see Ref.~\cite{Ota2023}). 
The beam size at the target position was checked by comparing the $^{11}$Be count rates with the phoswich during the two separate runs with and without the target holder. 
The difference in the count rates was negligible ($<$3\%) and was dominated by the beam-intensity fluctuation.
Therefore almost all the $^{11}$Be beam flux passed the target within the 17 mm diameter, without being blocked by the target holder. 
An additional test was made by measuring the beam halo with some inner rings of the Si 1\&2 telescope without the target holder. 
The halo count rates measured with the Si telescope were less than 1--2\% of the $^{11}$Be count rates measured with the phoswich. 
Through GEANT4 simulations, it was confirmed that such count rates were possible only when the beam's divergence from the telescope for the beam diagnosis is negligibly small ($\sigma_{\theta}<0.3^\circ$), leading to the beam spot size estimated at the reaction target position to be $\sigma_{x,y}<3$~mm. 

 \begin{figure}[b]
        \centering
          \includegraphics[width=8.5cm]{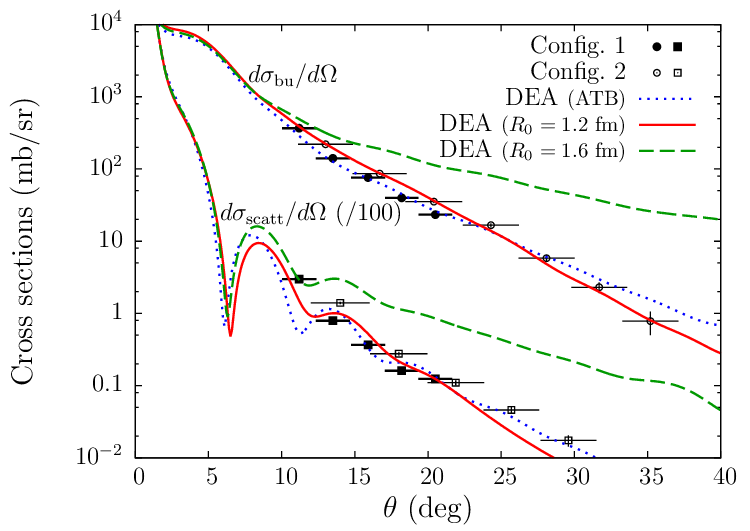}
        \caption{
Cross sections for scattering (squares; divided by 100) and inclusive breakup (circles) for $^{11}$Be on C at 22.8~MeV/nucleon as a function of the scattering angle of $^{11}$Be and $^{10}$Be in the center-of-mass frame, respectively.
Black and white symbols correspond to the data from Configs.\ 1 and 2, respectively.
Calculations have been performed within the DEA using different $U_{cT}$, see text for details.}\label{fig:Fig3}
\end{figure}

Figure~\ref{fig:Fig3} displays the measured angular differential cross sections for the scattering (squares; divided by 100 for readability) and inclusive breakup (circles) as a function of the scattering angle of $^{11}$Be and $^{10}$Be in the center-of-mass rest frame $\theta$, respectively.
The former cross section includes both elastic and inelastic events, since the energy resolution is not fine enough to exclude the 320~keV excitation energy of the first and only bound excited state of $^{11}$Be. 
However for the ratio method, the inelastic-scattering cross section does not need to be measured independently \cite{CJN11,CJN13}.
Moreover, our calculations show that it is at least an order of magnitude smaller than the elastic one \cite{Ota2023}.

The uncertainties (1 $\sigma$) of the measured cross sections are about 7--19\%, depending on the angle.
These uncertainties include statistical (1--17\%), beam intensity ($<$0.1\%), target thickness (3.5\%), and solid angle (5.4\%). 
The measured cross sections do not involve any data acquisition (DAQ) deadtime as both phoswich and Si arrays were run in the same DAQ system, cancelling each other's deadtime. 
The phoswich's efficiency for heavy ion detection was confirmed to be 100.0 (14)\% 
by comparing Li and light ion punch-through events from the Si 3\&4 telescope with those in coincidence with the phoswich.
This also ensures the coincidence efficiency between the phoswich and the Si telescopes.
The coincidence efficiency for $^{11,10}$Be between Si $\Delta E$-$E$ detectors were confirmed to be 100 (1)\% from the ratio between the coincidence-singles ratio of the $E$ detector at the corresponding energy region ($\sim100$ MeV). 
The uncertainty on the total number of $^{11}$Be beam particles was estimated by using different PID cuts in Fig.~\ref{fig:Fig2}(b). 
Nearly 97\% of the $^{11}$Be events are located in the peak position, which was used to determine the cross sections. 
This leads to less than 3\% underestimation of the 
 total number of beam particles.
This systematic uncertainty is not accounted for in the aforementioned estimate.
However, this is negligible for the total uncertainty. 

The experimental scattering and breakup cross sections shown in \Fig{fig:Fig3} exhibit very similar angular dependences, confirming the prediction that halo nuclei are scattered similarly whether they remain bound or break up \cite{CHB10}.
Taking their ratio should remove that global angular dependence and give us access to much more precise structure information about the projectile \cite{CJN11,CJN13}.
For both observables, Config.\ 1 leads systematically to a lower cross section than Config.\ 2 in the overlapping region.
This probably arises from an underestimation of the detection efficiency at angles shadowed by Si 1\&2 in Config.\ 1.
The ratio, however, should not depend on that systematic experimental uncertainty.

To analyze these data in more detail, we use the dynamical eikonal approximation (DEA) \cite{BCG05,GBC06}, which provides excellent results for the elastic scattering and diffractive breakup of one-nucleon halo nuclei on both light and heavy targets.
To account for the rather low beam energy of this experiment, we consider the correction analyzed in Ref.~\cite{FOC14}.
The DEA is built on the usual three-body model of reaction.
The projectile $P$ is described as a core $c$ to which a neutron $n$ is loosely bound.
The $c$-$n$ interaction is simulated by an effective potential $V_{cn}$.
Because it reproduces the predictions of the \emph{ab initio} calculation of $^{11}$Be by Calci \etal\  \cite{Cal16} and leads to excellent agreement with various breakup \cite{CPH18,MC19}, transfer \cite{YC18}, and knockout \cite{HC21} data, we use the Halo EFT interaction beyond next-to-leading order with the regulator $\sigma=1.2$~fm of Ref.~\cite{CPH18}.
It describes $^{11}$Be as a valence neutron loosely bound to a $^{10}$Be core in its $0^+$ ground state.
The $\half^+$ ground state and $\half^-$ excited state of $^{11}$Be are described as $1s_{1/2}$ and $0p_{1/2}$ neutron orbitals, respectively.
The $\fial^+$ resonance is modeled in the $d_{5/2}$ partial wave.
In all other partial waves $V_{cn}=0$, in the spirit of Halo EFT \cite{BHvK02,HJP17}.
The interaction of the projectile constituents with the target $T$ are described by optical potentials $U_{cT}$ and $U_{nT}$.
For $U_{cT}$, we first consider the phenomenological optical potential developed by Al-Khalili, Tostevin and Brooke (ATB) to reproduce the elastic scattering of $^{10}$Be off $^{12}$C at 59.4~MeV/nucleon \cite{ATB97}.
To estimate the sensitivity of our calculations to $U_{cT}$, we also use two double-folding potentials (DFPs) built from chiral EFT nucleon-nucleon interactions at N$^2$LO with cutoffs $R_0=1.2$ and 1.6~fm \cite{Dur18,DCS20}.
For $U_{nT}$, we adopt the global optical potential of Koning and Delaroche \cite{KD03}.

To obtain the inclusive breakup cross section,
we compute the reaction without the imaginary part of $U_{nT}$.
This way, the stripping events, in which the neutron is absorbed by the target, feed the diffractive-breakup channel.
Accordingly, the DEA breakup cross section is directly comparable to our inclusive measurement.
The scattering angle of the $^{10}$Be core after breakup is inferred from that of the $^{10}$Be-$n$ center of mass by assuming the neutron to be scattered at forward angles, an approximation supported by the data of Ref.~\cite{Anne94}.
Using a three-body effective interaction \cite{Capel2022}, we have estimated the contribution of the core excitation to the cross sections to be smaller than the experimental uncertainty.

The results of these calculations are shown in \Fig{fig:Fig3} alongside the data.
As expected for a collision of light nuclei, they strongly depend on the choice of optical potential \cite{CGB04}.
The phenomenological potential ATB (blue dotted lines) and the DFP with cutoff $R_0=1.2$~fm (solid red lines) provide a good agreement with the data for both the scattering and breakup angular distributions.
The DFP obtained with $R_0=1.6$~fm predicts too high cross sections for both observables.

\begin{figure}[!ht]
        \centering
          \includegraphics[width=8.5cm]{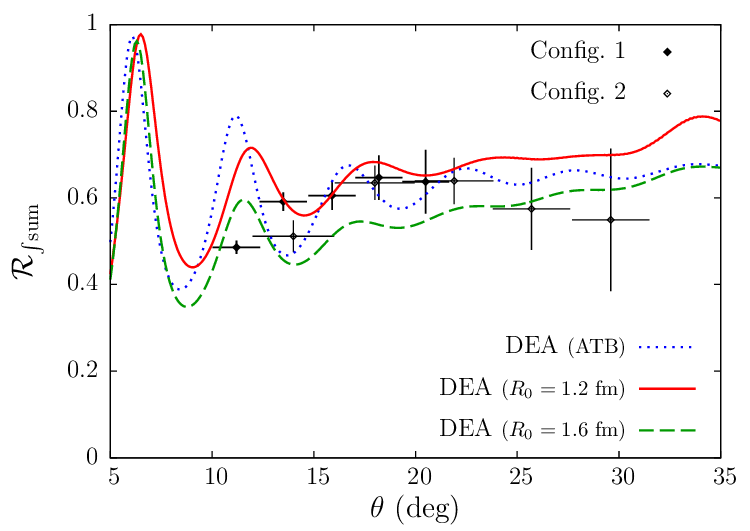}
        \caption{Ratio ${\cal R}_{\int\!\rm sum}$ \eq{e1} derived from the data and calculations of \Fig{fig:Fig3} as a function of the scattering angle $\theta$ of the $^{10}$Be-$n$ center of mass.}\label{fig:Fig4}
\end{figure}

From these data, we can compute the ratio ${\cal R}_{\int\!\rm sum}$ (see Eq.~(A3) of Ref.~\cite{CJN13})
\beq
{\cal R}_{\int\!\rm sum}&=&\frac{\int d\sigma_{\rm bu}/dEd\Omega\ dE}{d\sigma_{\rm scatt}/d\Omega+\int d\sigma_{\rm bu}/dEd\Omega\ dE},
\eeqn{e1}
which is the ratio of the breakup angular cross section integrated over all $c$-$n$ relative energies $E$,
divided by the sum of scattering and total breakup cross sections.
In the original idea, the ratio is expressed as a function of the $^{10}$Be-$n$ scattering angle, which we reconstruct from the measured angle of $^{10}$Be by assuming that $n$ is scattered at $0^\circ$.
This ratio, shown in \Fig{fig:Fig4},
 agrees very well with the calculations, independently of $U_{cT}$.
For instance, at $25^\circ$ the ratios differ by only 15\% whereas the cross sections change by more than an order of magnitude. 
This is the first key advantage of the ratio: being independent of the reaction process, it is nearly insensitive to the choice of optical potential \cite{CJN11,CJN13,CJN20}.
In addition, the systematic difference between the cross sections measured in Configs.\ 1 and 2 observed in \Fig{fig:Fig3} are no longer visible in the ratio.
The second advantage of this new reaction observable is that it cancels all systematic uncertainties.

\begin{figure}[!t]
        \centering
          \includegraphics[width=8.5cm]{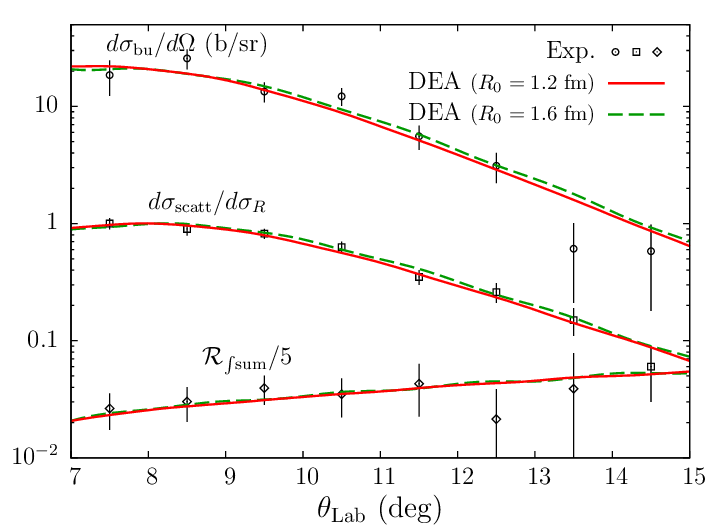}
        \caption{Ratio method for the collision of $^{11}$Be on Pb at 19.1~MeV/nucleon.
        DEA calculations are compared to the data of Ref.~\cite{Duan22} for the scattering (ratio to Rutherford) and inclusive breakup (in b/sr).
        The ratio inferred from the data is perfectly explained by theory, independently of $U_{cT}$.
        }\label{fig:Fig5}
\end{figure}

To extend our study to Coulomb-dominated reactions, we perform a similar analysis using the data for the collision of $^{11}$Be on Pb at 19.1~MeV/nucleon \cite{Duan22}.
In \Fig{fig:Fig5} we show the angular cross sections for inclusive breakup (in b/sr; circles), scattering (as a ratio to Rutherford; squares) and the ratio ${\cal R}_{\int\!\rm sum}$ (divided by 5 for readability; diamonds).
The DEA calculations have been performed with the same description of $^{11}$Be as on the carbon target.
The theoretical cross sections are in excellent agreement with the data using either DFP (solid red and dashed green lines) \cite{DC24}.
Accordingly, they reproduce very well the ratio inferred from the data.
Note that even the slight dependence upon the optical potential seen in each angular distribution is fully removed in the ratio.

The ratio has been suggested as a new reaction observable to study the structure of exotic nuclei \cite{CJN11,CJN13}.
The present study is the first firm experimental evidence that the method works for halo nuclei.
We show that it removes the strong angular dependence observed in the differential cross sections for breakup and scattering.
Moreover, our analysis confirms that the ratio is nearly independent of the optical-potential choice and does not need a precise normalization of the cross sections.

Theory suggests that the ratio method remains valid for non-halo nuclei \cite{CJN13,YCP19}.
It could therefore be used to study the single-particle structure of nuclei far from stability.
This poses various challenges.
Experimentally, it requires measuring the partial differential cross sections to each core state at large angles.
On the theory side, the core excitation has to be included explicitly in the reaction model \cite{ML12}.

The promising results obtained here will be further tested in the recently approved FRIB experiment, during which we will measure the diffractive breakup and scattering of the more exotic halo nucleus $^{19}$C up to large angles \cite{PAC3}.
By detecting the valence neutron in coincidence with the core, we will be able to measure accurately the core-$n$ scattering angle and relative energy.
These efforts will pave the way to establish the ratio method as a general spectroscopic tool for unstable nuclei.


We express our thanks to Dr.\ B.\ T.\ Roeder (Texas A\&M University) and the technical staff at the Texas A\&M University Cyclotron Institute. 
We thank E.\ A.\ Ricard from the Brookhaven National Laboratory for her constructive comments on this paper.
SO acknowledges support from the Office of Nuclear Physics, Office of Science of the U.S. Department of Energy under Contract No.\ DE-AC02-98CH10886 with Brookhaven Science Associates, LLC.\  
Financial support for this work was provided by the US Department of Energy, award Nos.\ DE-FG02-93ER40773 and DE-SC0018980, the US National Nuclear Security Administration, award No.\ DE-NA0003841. 
This project has received funding from the Deutsche Forschungsgemeinschaft within the Collaborative Research Center SFB 1245 (Projektnummer 279384907) and the PRISMA+ (Precision Physics, Fundamental Interactions and Structure of Matter) Cluster of Excellence.
GC acknowledges support from the National Sciences and Engineering Research Council of Canada, award SAPIN-2020-00052.
RCJ is supported by the UKRI/STFC research grant ST/V001108/1.
FMN is supported by the Department of Energy grant DE-SC0021422.
\bibliography{11Be_Manuscript240322}

\begin{thebibliography}{45}%
\makeatletter
\providecommand \@ifxundefined [1]{%
 \@ifx{#1\undefined}
}%
\providecommand \@ifnum [1]{%
 \ifnum #1\expandafter \@firstoftwo
 \else \expandafter \@secondoftwo
 \fi
}%
\providecommand \@ifx [1]{%
 \ifx #1\expandafter \@firstoftwo
 \else \expandafter \@secondoftwo
 \fi
}%
\providecommand \natexlab [1]{#1}%
\providecommand \enquote  [1]{``#1''}%
\providecommand \bibnamefont  [1]{#1}%
\providecommand \bibfnamefont [1]{#1}%
\providecommand \citenamefont [1]{#1}%
\providecommand \href@noop [0]{\@secondoftwo}%
\providecommand \href [0]{\begingroup \@sanitize@url \@href}%
\providecommand \@href[1]{\@@startlink{#1}\@@href}%
\providecommand \@@href[1]{\endgroup#1\@@endlink}%
\providecommand \@sanitize@url [0]{\catcode `\\12\catcode `\$12\catcode
  `\&12\catcode `\#12\catcode `\^12\catcode `\_12\catcode `\%12\relax}%
\providecommand \@@startlink[1]{}%
\providecommand \@@endlink[0]{}%
\providecommand \url  [0]{\begingroup\@sanitize@url \@url }%
\providecommand \@url [1]{\endgroup\@href {#1}{\urlprefix }}%
\providecommand \urlprefix  [0]{URL }%
\providecommand \Eprint [0]{\href }%
\providecommand \doibase [0]{http://dx.doi.org/}%
\providecommand \selectlanguage [0]{\@gobble}%
\providecommand \bibinfo  [0]{\@secondoftwo}%
\providecommand \bibfield  [0]{\@secondoftwo}%
\providecommand \translation [1]{[#1]}%
\providecommand \BibitemOpen [0]{}%
\providecommand \bibitemStop [0]{}%
\providecommand \bibitemNoStop [0]{.\EOS\space}%
\providecommand \EOS [0]{\spacefactor3000\relax}%
\providecommand \BibitemShut  [1]{\csname bibitem#1\endcsname}%
\let\auto@bib@innerbib\@empty
\bibitem [{\citenamefont {Tanihata}\ \emph
  {et~al.}(1985{\natexlab{a}})\citenamefont {Tanihata}, \citenamefont
  {Hamagaki}, \citenamefont {Hashimoto}, \citenamefont {Shida}, \citenamefont
  {Yoshikawa}, \citenamefont {Sugimoto}, \citenamefont {Yamakawa},
  \citenamefont {Kobayashi},\ and\ \citenamefont {Takahashi}}]{Tan85l}%
  \BibitemOpen
  \bibfield  {author} {\bibinfo {author} {\bibfnamefont {I.}~\bibnamefont
  {Tanihata}}, \bibinfo {author} {\bibfnamefont {H.}~\bibnamefont {Hamagaki}},
  \bibinfo {author} {\bibfnamefont {O.}~\bibnamefont {Hashimoto}}, \bibinfo
  {author} {\bibfnamefont {Y.}~\bibnamefont {Shida}}, \bibinfo {author}
  {\bibfnamefont {N.}~\bibnamefont {Yoshikawa}}, \bibinfo {author}
  {\bibfnamefont {K.}~\bibnamefont {Sugimoto}}, \bibinfo {author}
  {\bibfnamefont {O.}~\bibnamefont {Yamakawa}}, \bibinfo {author}
  {\bibfnamefont {T.}~\bibnamefont {Kobayashi}}, \ and\ \bibinfo {author}
  {\bibfnamefont {N.}~\bibnamefont {Takahashi}},\ }\href {\doibase
  10.1103/PhysRevLett.55.2676} {\bibfield  {journal} {\bibinfo  {journal}
  {Phys. Rev. Lett.}\ }\textbf {\bibinfo {volume} {55}},\ \bibinfo {pages}
  {2676} (\bibinfo {year} {1985}{\natexlab{a}})}\BibitemShut {NoStop}%
\bibitem [{\citenamefont {Tanihata}\ \emph
  {et~al.}(1985{\natexlab{b}})\citenamefont {Tanihata}, \citenamefont
  {Hamagaki}, \citenamefont {Hashimoto}, \citenamefont {Nagamiya},
  \citenamefont {Shida}, \citenamefont {Yoshikawa}, \citenamefont {Yamakawa},
  \citenamefont {Sugimoto}, \citenamefont {Kobayashi}, \citenamefont {Greiner},
  \citenamefont {Takahashi},\ and\ \citenamefont {Nojiri}}]{Tan85b}%
  \BibitemOpen
  \bibfield  {author} {\bibinfo {author} {\bibfnamefont {I.}~\bibnamefont
  {Tanihata}}, \bibinfo {author} {\bibfnamefont {H.}~\bibnamefont {Hamagaki}},
  \bibinfo {author} {\bibfnamefont {O.}~\bibnamefont {Hashimoto}}, \bibinfo
  {author} {\bibfnamefont {S.}~\bibnamefont {Nagamiya}}, \bibinfo {author}
  {\bibfnamefont {Y.}~\bibnamefont {Shida}}, \bibinfo {author} {\bibfnamefont
  {N.}~\bibnamefont {Yoshikawa}}, \bibinfo {author} {\bibfnamefont
  {O.}~\bibnamefont {Yamakawa}}, \bibinfo {author} {\bibfnamefont
  {K.}~\bibnamefont {Sugimoto}}, \bibinfo {author} {\bibfnamefont
  {T.}~\bibnamefont {Kobayashi}}, \bibinfo {author} {\bibfnamefont
  {D.}~\bibnamefont {Greiner}}, \bibinfo {author} {\bibfnamefont
  {N.}~\bibnamefont {Takahashi}}, \ and\ \bibinfo {author} {\bibfnamefont
  {Y.}~\bibnamefont {Nojiri}},\ }\href {\doibase
  https://doi.org/10.1016/0370-2693(85)90005-X} {\bibfield  {journal} {\bibinfo
   {journal} {Phys. Lett.}\ }\textbf {\bibinfo {volume} {B160}},\ \bibinfo
  {pages} {380} (\bibinfo {year} {1985}{\natexlab{b}})}\BibitemShut {NoStop}%
\bibitem [{\citenamefont {Hansen}\ and\ \citenamefont {Jonson}(1987)}]{HJ87}%
  \BibitemOpen
  \bibfield  {author} {\bibinfo {author} {\bibfnamefont {P.~G.}\ \bibnamefont
  {Hansen}}\ and\ \bibinfo {author} {\bibfnamefont {B.}~\bibnamefont
  {Jonson}},\ }\href {\doibase 10.1209/0295-5075/4/4/005} {\bibfield  {journal}
  {\bibinfo  {journal} {Europhys. Lett.}\ }\textbf {\bibinfo {volume} {4}},\
  \bibinfo {pages} {409} (\bibinfo {year} {1987})}\BibitemShut {NoStop}%
\bibitem [{\citenamefont {Tanihata}(1996)}]{Tan96}%
  \BibitemOpen
  \bibfield  {author} {\bibinfo {author} {\bibfnamefont {I.}~\bibnamefont
  {Tanihata}},\ }\href {\doibase 10.1088/0954-3899/22/2/004} {\bibfield
  {journal} {\bibinfo  {journal} {J. Phys. G}\ }\textbf {\bibinfo {volume}
  {22}},\ \bibinfo {pages} {157} (\bibinfo {year} {1996})}\BibitemShut
  {NoStop}%
\bibitem [{\citenamefont {Aumann}\ \emph {et~al.}(2000)\citenamefont {Aumann},
  \citenamefont {Navin}, \citenamefont {Balamuth}, \citenamefont {Bazin},
  \citenamefont {Blank}, \citenamefont {Brown}, \citenamefont {Bush},
  \citenamefont {Caggiano}, \citenamefont {Davids}, \citenamefont {Glasmacher},
  \citenamefont {Guimar\~aes}, \citenamefont {Hansen}, \citenamefont
  {Ibbotson}, \citenamefont {Karnes}, \citenamefont {Kolata}, \citenamefont
  {Maddalena}, \citenamefont {Pritychenko}, \citenamefont {Scheit},
  \citenamefont {Sherrill},\ and\ \citenamefont {Tostevin}}]{Aum00}%
  \BibitemOpen
  \bibfield  {author} {\bibinfo {author} {\bibfnamefont {T.}~\bibnamefont
  {Aumann}}, \bibinfo {author} {\bibfnamefont {A.}~\bibnamefont {Navin}},
  \bibinfo {author} {\bibfnamefont {D.~P.}\ \bibnamefont {Balamuth}}, \bibinfo
  {author} {\bibfnamefont {D.}~\bibnamefont {Bazin}}, \bibinfo {author}
  {\bibfnamefont {B.}~\bibnamefont {Blank}}, \bibinfo {author} {\bibfnamefont
  {B.~A.}\ \bibnamefont {Brown}}, \bibinfo {author} {\bibfnamefont {J.~E.}\
  \bibnamefont {Bush}}, \bibinfo {author} {\bibfnamefont {J.~A.}\ \bibnamefont
  {Caggiano}}, \bibinfo {author} {\bibfnamefont {B.}~\bibnamefont {Davids}},
  \bibinfo {author} {\bibfnamefont {T.}~\bibnamefont {Glasmacher}}, \bibinfo
  {author} {\bibfnamefont {V.}~\bibnamefont {Guimar\~aes}}, \bibinfo {author}
  {\bibfnamefont {P.~G.}\ \bibnamefont {Hansen}}, \bibinfo {author}
  {\bibfnamefont {R.~W.}\ \bibnamefont {Ibbotson}}, \bibinfo {author}
  {\bibfnamefont {D.}~\bibnamefont {Karnes}}, \bibinfo {author} {\bibfnamefont
  {J.~J.}\ \bibnamefont {Kolata}}, \bibinfo {author} {\bibfnamefont
  {V.}~\bibnamefont {Maddalena}}, \bibinfo {author} {\bibfnamefont
  {B.}~\bibnamefont {Pritychenko}}, \bibinfo {author} {\bibfnamefont
  {H.}~\bibnamefont {Scheit}}, \bibinfo {author} {\bibfnamefont {B.~M.}\
  \bibnamefont {Sherrill}}, \ and\ \bibinfo {author} {\bibfnamefont {J.~A.}\
  \bibnamefont {Tostevin}},\ }\href {\doibase 10.1103/PhysRevLett.84.35}
  {\bibfield  {journal} {\bibinfo  {journal} {Phys. Rev. Lett.}\ }\textbf
  {\bibinfo {volume} {84}},\ \bibinfo {pages} {35} (\bibinfo {year}
  {2000})}\BibitemShut {NoStop}%
\bibitem [{\citenamefont {Palit}\ \emph {et~al.}(2003)\citenamefont {Palit},
  \citenamefont {Adrich}, \citenamefont {Aumann}, \citenamefont {Boretzky},
  \citenamefont {Carlson}, \citenamefont {Cortina}, \citenamefont
  {Datta~Pramanik}, \citenamefont {Elze}, \citenamefont {Emling}, \citenamefont
  {Geissel}, \citenamefont {Hellstr\"om}, \citenamefont {Jones}, \citenamefont
  {Kratz}, \citenamefont {Kulessa}, \citenamefont {Leifels}, \citenamefont
  {Leistenschneider}, \citenamefont {M\"unzenberg}, \citenamefont {Nociforo},
  \citenamefont {Reiter}, \citenamefont {Simon}, \citenamefont {S\"ummerer},\
  and\ \citenamefont {Walus}}]{Pal03}%
  \BibitemOpen
  \bibfield  {author} {\bibinfo {author} {\bibfnamefont {R.}~\bibnamefont
  {Palit}}, \bibinfo {author} {\bibfnamefont {P.}~\bibnamefont {Adrich}},
  \bibinfo {author} {\bibfnamefont {T.}~\bibnamefont {Aumann}}, \bibinfo
  {author} {\bibfnamefont {K.}~\bibnamefont {Boretzky}}, \bibinfo {author}
  {\bibfnamefont {B.~V.}\ \bibnamefont {Carlson}}, \bibinfo {author}
  {\bibfnamefont {D.}~\bibnamefont {Cortina}}, \bibinfo {author} {\bibfnamefont
  {U.}~\bibnamefont {Datta~Pramanik}}, \bibinfo {author} {\bibfnamefont
  {T.~W.}\ \bibnamefont {Elze}}, \bibinfo {author} {\bibfnamefont
  {H.}~\bibnamefont {Emling}}, \bibinfo {author} {\bibfnamefont
  {H.}~\bibnamefont {Geissel}}, \bibinfo {author} {\bibfnamefont
  {M.}~\bibnamefont {Hellstr\"om}}, \bibinfo {author} {\bibfnamefont {K.~L.}\
  \bibnamefont {Jones}}, \bibinfo {author} {\bibfnamefont {J.~V.}\ \bibnamefont
  {Kratz}}, \bibinfo {author} {\bibfnamefont {R.}~\bibnamefont {Kulessa}},
  \bibinfo {author} {\bibfnamefont {Y.}~\bibnamefont {Leifels}}, \bibinfo
  {author} {\bibfnamefont {A.}~\bibnamefont {Leistenschneider}}, \bibinfo
  {author} {\bibfnamefont {G.}~\bibnamefont {M\"unzenberg}}, \bibinfo {author}
  {\bibfnamefont {C.}~\bibnamefont {Nociforo}}, \bibinfo {author}
  {\bibfnamefont {P.}~\bibnamefont {Reiter}}, \bibinfo {author} {\bibfnamefont
  {H.}~\bibnamefont {Simon}}, \bibinfo {author} {\bibfnamefont
  {K.}~\bibnamefont {S\"ummerer}}, \ and\ \bibinfo {author} {\bibfnamefont
  {W.}~\bibnamefont {Walus}} (\bibinfo {collaboration} {LAND/FRS
  Collaboration}),\ }\href {\doibase 10.1103/PhysRevC.68.034318} {\bibfield
  {journal} {\bibinfo  {journal} {Phys. Rev. C}\ }\textbf {\bibinfo {volume}
  {68}},\ \bibinfo {pages} {034318} (\bibinfo {year} {2003})}\BibitemShut
  {NoStop}%
\bibitem [{\citenamefont {Fukuda}\ \emph {et~al.}(2004)\citenamefont {Fukuda},
  \citenamefont {Nakamura}, \citenamefont {Aoi}, \citenamefont {Imai},
  \citenamefont {Ishihara}, \citenamefont {Kobayashi}, \citenamefont {Iwasaki},
  \citenamefont {Kubo}, \citenamefont {Mengoni}, \citenamefont {Notani},
  \citenamefont {Otsu}, \citenamefont {Sakurai}, \citenamefont {Shimoura},
  \citenamefont {Teranishi}, \citenamefont {Watanabe},\ and\ \citenamefont
  {Yoneda}}]{Fuk04}%
  \BibitemOpen
  \bibfield  {author} {\bibinfo {author} {\bibfnamefont {N.}~\bibnamefont
  {Fukuda}}, \bibinfo {author} {\bibfnamefont {T.}~\bibnamefont {Nakamura}},
  \bibinfo {author} {\bibfnamefont {N.}~\bibnamefont {Aoi}}, \bibinfo {author}
  {\bibfnamefont {N.}~\bibnamefont {Imai}}, \bibinfo {author} {\bibfnamefont
  {M.}~\bibnamefont {Ishihara}}, \bibinfo {author} {\bibfnamefont
  {T.}~\bibnamefont {Kobayashi}}, \bibinfo {author} {\bibfnamefont
  {H.}~\bibnamefont {Iwasaki}}, \bibinfo {author} {\bibfnamefont
  {T.}~\bibnamefont {Kubo}}, \bibinfo {author} {\bibfnamefont {A.}~\bibnamefont
  {Mengoni}}, \bibinfo {author} {\bibfnamefont {M.}~\bibnamefont {Notani}},
  \bibinfo {author} {\bibfnamefont {H.}~\bibnamefont {Otsu}}, \bibinfo {author}
  {\bibfnamefont {H.}~\bibnamefont {Sakurai}}, \bibinfo {author} {\bibfnamefont
  {S.}~\bibnamefont {Shimoura}}, \bibinfo {author} {\bibfnamefont
  {T.}~\bibnamefont {Teranishi}}, \bibinfo {author} {\bibfnamefont {Y.~X.}\
  \bibnamefont {Watanabe}}, \ and\ \bibinfo {author} {\bibfnamefont
  {K.}~\bibnamefont {Yoneda}},\ }\href {\doibase 10.1103/PhysRevC.70.054606}
  {\bibfield  {journal} {\bibinfo  {journal} {Phys. Rev. C}\ }\textbf {\bibinfo
  {volume} {70}},\ \bibinfo {pages} {054606} (\bibinfo {year}
  {2004})}\BibitemShut {NoStop}%
\bibitem [{\citenamefont {Di~Pietro}\ \emph {et~al.}(2010)\citenamefont
  {Di~Pietro}, \citenamefont {Randisi}, \citenamefont {Scuderi}, \citenamefont
  {Acosta}, \citenamefont {Amorini}, \citenamefont {Borge}, \citenamefont
  {Figuera}, \citenamefont {Fisichella}, \citenamefont {Fraile}, \citenamefont
  {Gomez-Camacho}, \citenamefont {Jeppesen}, \citenamefont {Lattuada},
  \citenamefont {Martel}, \citenamefont {Milin}, \citenamefont {Musumarra},
  \citenamefont {Papa}, \citenamefont {Pellegriti}, \citenamefont
  {Perez-Bernal}, \citenamefont {Raabe}, \citenamefont {Rizzo}, \citenamefont
  {Santonocito}, \citenamefont {Scalia}, \citenamefont {Tengblad},
  \citenamefont {Torresi}, \citenamefont {Vidal}, \citenamefont {Voulot},
  \citenamefont {Wenander},\ and\ \citenamefont {Zadro}}]{DiP10}%
  \BibitemOpen
  \bibfield  {author} {\bibinfo {author} {\bibfnamefont {A.}~\bibnamefont
  {Di~Pietro}}, \bibinfo {author} {\bibfnamefont {G.}~\bibnamefont {Randisi}},
  \bibinfo {author} {\bibfnamefont {V.}~\bibnamefont {Scuderi}}, \bibinfo
  {author} {\bibfnamefont {L.}~\bibnamefont {Acosta}}, \bibinfo {author}
  {\bibfnamefont {F.}~\bibnamefont {Amorini}}, \bibinfo {author} {\bibfnamefont
  {M.~J.~G.}\ \bibnamefont {Borge}}, \bibinfo {author} {\bibfnamefont
  {P.}~\bibnamefont {Figuera}}, \bibinfo {author} {\bibfnamefont
  {M.}~\bibnamefont {Fisichella}}, \bibinfo {author} {\bibfnamefont {L.~M.}\
  \bibnamefont {Fraile}}, \bibinfo {author} {\bibfnamefont {J.}~\bibnamefont
  {Gomez-Camacho}}, \bibinfo {author} {\bibfnamefont {H.}~\bibnamefont
  {Jeppesen}}, \bibinfo {author} {\bibfnamefont {M.}~\bibnamefont {Lattuada}},
  \bibinfo {author} {\bibfnamefont {I.}~\bibnamefont {Martel}}, \bibinfo
  {author} {\bibfnamefont {M.}~\bibnamefont {Milin}}, \bibinfo {author}
  {\bibfnamefont {A.}~\bibnamefont {Musumarra}}, \bibinfo {author}
  {\bibfnamefont {M.}~\bibnamefont {Papa}}, \bibinfo {author} {\bibfnamefont
  {M.~G.}\ \bibnamefont {Pellegriti}}, \bibinfo {author} {\bibfnamefont
  {F.}~\bibnamefont {Perez-Bernal}}, \bibinfo {author} {\bibfnamefont
  {R.}~\bibnamefont {Raabe}}, \bibinfo {author} {\bibfnamefont
  {F.}~\bibnamefont {Rizzo}}, \bibinfo {author} {\bibfnamefont
  {D.}~\bibnamefont {Santonocito}}, \bibinfo {author} {\bibfnamefont
  {G.}~\bibnamefont {Scalia}}, \bibinfo {author} {\bibfnamefont
  {O.}~\bibnamefont {Tengblad}}, \bibinfo {author} {\bibfnamefont
  {D.}~\bibnamefont {Torresi}}, \bibinfo {author} {\bibfnamefont {A.~M.}\
  \bibnamefont {Vidal}}, \bibinfo {author} {\bibfnamefont {D.}~\bibnamefont
  {Voulot}}, \bibinfo {author} {\bibfnamefont {F.}~\bibnamefont {Wenander}}, \
  and\ \bibinfo {author} {\bibfnamefont {M.}~\bibnamefont {Zadro}},\ }\href
  {\doibase 10.1103/PhysRevLett.105.022701} {\bibfield  {journal} {\bibinfo
  {journal} {Phys. Rev. Lett.}\ }\textbf {\bibinfo {volume} {105}},\ \bibinfo
  {pages} {022701} (\bibinfo {year} {2010})}\BibitemShut {NoStop}%
\bibitem [{\citenamefont {Di~Pietro}\ \emph {et~al.}(2012)\citenamefont
  {Di~Pietro}, \citenamefont {Scuderi}, \citenamefont {Moro}, \citenamefont
  {Acosta}, \citenamefont {Amorini}, \citenamefont {Borge}, \citenamefont
  {Figuera}, \citenamefont {Fisichella}, \citenamefont {Fraile}, \citenamefont
  {Gomez-Camacho}, \citenamefont {Jeppesen}, \citenamefont {Lattuada},
  \citenamefont {Martel}, \citenamefont {Milin}, \citenamefont {Musumarra},
  \citenamefont {Papa}, \citenamefont {Pellegriti}, \citenamefont
  {Perez-Bernal}, \citenamefont {Raabe}, \citenamefont {Randisi}, \citenamefont
  {Rizzo}, \citenamefont {Scalia}, \citenamefont {Tengblad}, \citenamefont
  {Torresi}, \citenamefont {Vidal}, \citenamefont {Voulot}, \citenamefont
  {Wenander},\ and\ \citenamefont {Zadro}}]{DiP12}%
  \BibitemOpen
  \bibfield  {author} {\bibinfo {author} {\bibfnamefont {A.}~\bibnamefont
  {Di~Pietro}}, \bibinfo {author} {\bibfnamefont {V.}~\bibnamefont {Scuderi}},
  \bibinfo {author} {\bibfnamefont {A.~M.}\ \bibnamefont {Moro}}, \bibinfo
  {author} {\bibfnamefont {L.}~\bibnamefont {Acosta}}, \bibinfo {author}
  {\bibfnamefont {F.}~\bibnamefont {Amorini}}, \bibinfo {author} {\bibfnamefont
  {M.~J.~G.}\ \bibnamefont {Borge}}, \bibinfo {author} {\bibfnamefont
  {P.}~\bibnamefont {Figuera}}, \bibinfo {author} {\bibfnamefont
  {M.}~\bibnamefont {Fisichella}}, \bibinfo {author} {\bibfnamefont {L.~M.}\
  \bibnamefont {Fraile}}, \bibinfo {author} {\bibfnamefont {J.}~\bibnamefont
  {Gomez-Camacho}}, \bibinfo {author} {\bibfnamefont {H.}~\bibnamefont
  {Jeppesen}}, \bibinfo {author} {\bibfnamefont {M.}~\bibnamefont {Lattuada}},
  \bibinfo {author} {\bibfnamefont {I.}~\bibnamefont {Martel}}, \bibinfo
  {author} {\bibfnamefont {M.}~\bibnamefont {Milin}}, \bibinfo {author}
  {\bibfnamefont {A.}~\bibnamefont {Musumarra}}, \bibinfo {author}
  {\bibfnamefont {M.}~\bibnamefont {Papa}}, \bibinfo {author} {\bibfnamefont
  {M.~G.}\ \bibnamefont {Pellegriti}}, \bibinfo {author} {\bibfnamefont
  {F.}~\bibnamefont {Perez-Bernal}}, \bibinfo {author} {\bibfnamefont
  {R.}~\bibnamefont {Raabe}}, \bibinfo {author} {\bibfnamefont
  {G.}~\bibnamefont {Randisi}}, \bibinfo {author} {\bibfnamefont
  {F.}~\bibnamefont {Rizzo}}, \bibinfo {author} {\bibfnamefont
  {G.}~\bibnamefont {Scalia}}, \bibinfo {author} {\bibfnamefont
  {O.}~\bibnamefont {Tengblad}}, \bibinfo {author} {\bibfnamefont
  {D.}~\bibnamefont {Torresi}}, \bibinfo {author} {\bibfnamefont {A.~M.}\
  \bibnamefont {Vidal}}, \bibinfo {author} {\bibfnamefont {D.}~\bibnamefont
  {Voulot}}, \bibinfo {author} {\bibfnamefont {F.}~\bibnamefont {Wenander}}, \
  and\ \bibinfo {author} {\bibfnamefont {M.}~\bibnamefont {Zadro}},\ }\href
  {\doibase 10.1103/PhysRevC.85.054607} {\bibfield  {journal} {\bibinfo
  {journal} {Phys. Rev. C}\ }\textbf {\bibinfo {volume} {85}},\ \bibinfo
  {pages} {054607} (\bibinfo {year} {2012})}\BibitemShut {NoStop}%
\bibitem [{\citenamefont {Schmitt}\ \emph {et~al.}(2012)\citenamefont
  {Schmitt}, \citenamefont {Jones}, \citenamefont {Bey}, \citenamefont {Ahn},
  \citenamefont {Bardayan}, \citenamefont {Blackmon}, \citenamefont {Brown},
  \citenamefont {Chae}, \citenamefont {Chipps}, \citenamefont {Cizewski},
  \citenamefont {Hahn}, \citenamefont {Kolata}, \citenamefont {Kozub},
  \citenamefont {Liang}, \citenamefont {Matei}, \citenamefont
  {Mato\ifmmode~\check{s}\else \v{s}\fi{}}, \citenamefont {Matyas},
  \citenamefont {Moazen}, \citenamefont {Nesaraja}, \citenamefont {Nunes},
  \citenamefont {O'Malley}, \citenamefont {Pain}, \citenamefont {Peters},
  \citenamefont {Pittman}, \citenamefont {Roberts}, \citenamefont {Shapira},
  \citenamefont {Shriner}, \citenamefont {Smith}, \citenamefont {Spassova},
  \citenamefont {Stracener}, \citenamefont {Villano},\ and\ \citenamefont
  {Wilson}}]{Sch12}%
  \BibitemOpen
  \bibfield  {author} {\bibinfo {author} {\bibfnamefont {K.~T.}\ \bibnamefont
  {Schmitt}}, \bibinfo {author} {\bibfnamefont {K.~L.}\ \bibnamefont {Jones}},
  \bibinfo {author} {\bibfnamefont {A.}~\bibnamefont {Bey}}, \bibinfo {author}
  {\bibfnamefont {S.~H.}\ \bibnamefont {Ahn}}, \bibinfo {author} {\bibfnamefont
  {D.~W.}\ \bibnamefont {Bardayan}}, \bibinfo {author} {\bibfnamefont {J.~C.}\
  \bibnamefont {Blackmon}}, \bibinfo {author} {\bibfnamefont {S.~M.}\
  \bibnamefont {Brown}}, \bibinfo {author} {\bibfnamefont {K.~Y.}\ \bibnamefont
  {Chae}}, \bibinfo {author} {\bibfnamefont {K.~A.}\ \bibnamefont {Chipps}},
  \bibinfo {author} {\bibfnamefont {J.~A.}\ \bibnamefont {Cizewski}}, \bibinfo
  {author} {\bibfnamefont {K.~I.}\ \bibnamefont {Hahn}}, \bibinfo {author}
  {\bibfnamefont {J.~J.}\ \bibnamefont {Kolata}}, \bibinfo {author}
  {\bibfnamefont {R.~L.}\ \bibnamefont {Kozub}}, \bibinfo {author}
  {\bibfnamefont {J.~F.}\ \bibnamefont {Liang}}, \bibinfo {author}
  {\bibfnamefont {C.}~\bibnamefont {Matei}}, \bibinfo {author} {\bibfnamefont
  {M.}~\bibnamefont {Mato\ifmmode~\check{s}\else \v{s}\fi{}}}, \bibinfo
  {author} {\bibfnamefont {D.}~\bibnamefont {Matyas}}, \bibinfo {author}
  {\bibfnamefont {B.}~\bibnamefont {Moazen}}, \bibinfo {author} {\bibfnamefont
  {C.}~\bibnamefont {Nesaraja}}, \bibinfo {author} {\bibfnamefont {F.~M.}\
  \bibnamefont {Nunes}}, \bibinfo {author} {\bibfnamefont {P.~D.}\ \bibnamefont
  {O'Malley}}, \bibinfo {author} {\bibfnamefont {S.~D.}\ \bibnamefont {Pain}},
  \bibinfo {author} {\bibfnamefont {W.~A.}\ \bibnamefont {Peters}}, \bibinfo
  {author} {\bibfnamefont {S.~T.}\ \bibnamefont {Pittman}}, \bibinfo {author}
  {\bibfnamefont {A.}~\bibnamefont {Roberts}}, \bibinfo {author} {\bibfnamefont
  {D.}~\bibnamefont {Shapira}}, \bibinfo {author} {\bibfnamefont {J.~F.}\
  \bibnamefont {Shriner}}, \bibinfo {author} {\bibfnamefont {M.~S.}\
  \bibnamefont {Smith}}, \bibinfo {author} {\bibfnamefont {I.}~\bibnamefont
  {Spassova}}, \bibinfo {author} {\bibfnamefont {D.~W.}\ \bibnamefont
  {Stracener}}, \bibinfo {author} {\bibfnamefont {A.~N.}\ \bibnamefont
  {Villano}}, \ and\ \bibinfo {author} {\bibfnamefont {G.~L.}\ \bibnamefont
  {Wilson}},\ }\href {\doibase 10.1103/PhysRevLett.108.192701} {\bibfield
  {journal} {\bibinfo  {journal} {Phys. Rev. Lett.}\ }\textbf {\bibinfo
  {volume} {108}},\ \bibinfo {pages} {192701} (\bibinfo {year}
  {2012})}\BibitemShut {NoStop}%
\bibitem [{\citenamefont {Schmitt}\ \emph {et~al.}(2013)\citenamefont
  {Schmitt}, \citenamefont {Jones}, \citenamefont {Ahn}, \citenamefont
  {Bardayan}, \citenamefont {Bey}, \citenamefont {Blackmon}, \citenamefont
  {Brown}, \citenamefont {Chae}, \citenamefont {Chipps}, \citenamefont
  {Cizewski}, \citenamefont {Hahn}, \citenamefont {Kolata}, \citenamefont
  {Kozub}, \citenamefont {Liang}, \citenamefont {Matei}, \citenamefont {Matos},
  \citenamefont {Matyas}, \citenamefont {Moazen}, \citenamefont {Nesaraja},
  \citenamefont {Nunes}, \citenamefont {O'Malley}, \citenamefont {Pain},
  \citenamefont {Peters}, \citenamefont {Pittman}, \citenamefont {Roberts},
  \citenamefont {Shapira}, \citenamefont {Shriner}, \citenamefont {Smith},
  \citenamefont {Spassova}, \citenamefont {Stracener}, \citenamefont
  {Upadhyay}, \citenamefont {Villano},\ and\ \citenamefont {Wilson}}]{Sch13}%
  \BibitemOpen
  \bibfield  {author} {\bibinfo {author} {\bibfnamefont {K.~T.}\ \bibnamefont
  {Schmitt}}, \bibinfo {author} {\bibfnamefont {K.~L.}\ \bibnamefont {Jones}},
  \bibinfo {author} {\bibfnamefont {S.}~\bibnamefont {Ahn}}, \bibinfo {author}
  {\bibfnamefont {D.~W.}\ \bibnamefont {Bardayan}}, \bibinfo {author}
  {\bibfnamefont {A.}~\bibnamefont {Bey}}, \bibinfo {author} {\bibfnamefont
  {J.~C.}\ \bibnamefont {Blackmon}}, \bibinfo {author} {\bibfnamefont {S.~M.}\
  \bibnamefont {Brown}}, \bibinfo {author} {\bibfnamefont {K.~Y.}\ \bibnamefont
  {Chae}}, \bibinfo {author} {\bibfnamefont {K.~A.}\ \bibnamefont {Chipps}},
  \bibinfo {author} {\bibfnamefont {J.~A.}\ \bibnamefont {Cizewski}}, \bibinfo
  {author} {\bibfnamefont {K.~I.}\ \bibnamefont {Hahn}}, \bibinfo {author}
  {\bibfnamefont {J.~J.}\ \bibnamefont {Kolata}}, \bibinfo {author}
  {\bibfnamefont {R.~L.}\ \bibnamefont {Kozub}}, \bibinfo {author}
  {\bibfnamefont {J.~F.}\ \bibnamefont {Liang}}, \bibinfo {author}
  {\bibfnamefont {C.}~\bibnamefont {Matei}}, \bibinfo {author} {\bibfnamefont
  {M.}~\bibnamefont {Matos}}, \bibinfo {author} {\bibfnamefont
  {D.}~\bibnamefont {Matyas}}, \bibinfo {author} {\bibfnamefont
  {B.}~\bibnamefont {Moazen}}, \bibinfo {author} {\bibfnamefont {C.~D.}\
  \bibnamefont {Nesaraja}}, \bibinfo {author} {\bibfnamefont {F.~M.}\
  \bibnamefont {Nunes}}, \bibinfo {author} {\bibfnamefont {P.~D.}\ \bibnamefont
  {O'Malley}}, \bibinfo {author} {\bibfnamefont {S.~D.}\ \bibnamefont {Pain}},
  \bibinfo {author} {\bibfnamefont {W.~A.}\ \bibnamefont {Peters}}, \bibinfo
  {author} {\bibfnamefont {S.~T.}\ \bibnamefont {Pittman}}, \bibinfo {author}
  {\bibfnamefont {A.}~\bibnamefont {Roberts}}, \bibinfo {author} {\bibfnamefont
  {D.}~\bibnamefont {Shapira}}, \bibinfo {author} {\bibfnamefont {J.~F.}\
  \bibnamefont {Shriner}}, \bibinfo {author} {\bibfnamefont {M.~S.}\
  \bibnamefont {Smith}}, \bibinfo {author} {\bibfnamefont {I.}~\bibnamefont
  {Spassova}}, \bibinfo {author} {\bibfnamefont {D.~W.}\ \bibnamefont
  {Stracener}}, \bibinfo {author} {\bibfnamefont {N.~J.}\ \bibnamefont
  {Upadhyay}}, \bibinfo {author} {\bibfnamefont {A.~N.}\ \bibnamefont
  {Villano}}, \ and\ \bibinfo {author} {\bibfnamefont {G.~L.}\ \bibnamefont
  {Wilson}},\ }\href {\doibase 10.1103/PhysRevC.88.064612} {\bibfield
  {journal} {\bibinfo  {journal} {Phys. Rev. C}\ }\textbf {\bibinfo {volume}
  {88}},\ \bibinfo {pages} {064612} (\bibinfo {year} {2013})}\BibitemShut
  {NoStop}%
\bibitem [{\citenamefont {Capel}\ \emph {et~al.}(2004)\citenamefont {Capel},
  \citenamefont {Goldstein},\ and\ \citenamefont {Baye}}]{CGB04}%
  \BibitemOpen
  \bibfield  {author} {\bibinfo {author} {\bibfnamefont {P.}~\bibnamefont
  {Capel}}, \bibinfo {author} {\bibfnamefont {G.}~\bibnamefont {Goldstein}}, \
  and\ \bibinfo {author} {\bibfnamefont {D.}~\bibnamefont {Baye}},\ }\href
  {\doibase 10.1103/PhysRevC.70.064605} {\bibfield  {journal} {\bibinfo
  {journal} {Phys. Rev. C}\ }\textbf {\bibinfo {volume} {70}},\ \bibinfo
  {pages} {064605} (\bibinfo {year} {2004})}\BibitemShut {NoStop}%
\bibitem [{\citenamefont {S\"urer}\ \emph {et~al.}(2022)\citenamefont
  {S\"urer}, \citenamefont {Nunes}, \citenamefont {Plumlee},\ and\
  \citenamefont {Wild}}]{SNP22}%
  \BibitemOpen
  \bibfield  {author} {\bibinfo {author} {\bibfnamefont {O.}~\bibnamefont
  {S\"urer}}, \bibinfo {author} {\bibfnamefont {F.~M.}\ \bibnamefont {Nunes}},
  \bibinfo {author} {\bibfnamefont {M.}~\bibnamefont {Plumlee}}, \ and\
  \bibinfo {author} {\bibfnamefont {S.~M.}\ \bibnamefont {Wild}},\ }\href
  {\doibase 10.1103/PhysRevC.106.024607} {\bibfield  {journal} {\bibinfo
  {journal} {Phys. Rev. C}\ }\textbf {\bibinfo {volume} {106}},\ \bibinfo
  {pages} {024607} (\bibinfo {year} {2022})}\BibitemShut {NoStop}%
\bibitem [{\citenamefont {Hebborn}\ \emph {et~al.}(2023)\citenamefont
  {Hebborn}, \citenamefont {Whitehead}, \citenamefont {Lovell},\ and\
  \citenamefont {Nunes}}]{HWL23}%
  \BibitemOpen
  \bibfield  {author} {\bibinfo {author} {\bibfnamefont {C.}~\bibnamefont
  {Hebborn}}, \bibinfo {author} {\bibfnamefont {T.~R.}\ \bibnamefont
  {Whitehead}}, \bibinfo {author} {\bibfnamefont {A.~E.}\ \bibnamefont
  {Lovell}}, \ and\ \bibinfo {author} {\bibfnamefont {F.~M.}\ \bibnamefont
  {Nunes}},\ }\href {\doibase 10.1103/PhysRevC.108.014601} {\bibfield
  {journal} {\bibinfo  {journal} {Phys. Rev. C}\ }\textbf {\bibinfo {volume}
  {108}},\ \bibinfo {pages} {014601} (\bibinfo {year} {2023})}\BibitemShut
  {NoStop}%
\bibitem [{\citenamefont {Catacora-Rios}\ \emph {et~al.}(2023)\citenamefont
  {Catacora-Rios}, \citenamefont {Lovell},\ and\ \citenamefont
  {Nunes}}]{CLN23}%
  \BibitemOpen
  \bibfield  {author} {\bibinfo {author} {\bibfnamefont {M.}~\bibnamefont
  {Catacora-Rios}}, \bibinfo {author} {\bibfnamefont {A.~E.}\ \bibnamefont
  {Lovell}}, \ and\ \bibinfo {author} {\bibfnamefont {F.~M.}\ \bibnamefont
  {Nunes}},\ }\href {\doibase 10.1103/PhysRevC.108.024601} {\bibfield
  {journal} {\bibinfo  {journal} {Phys. Rev. C}\ }\textbf {\bibinfo {volume}
  {108}},\ \bibinfo {pages} {024601} (\bibinfo {year} {2023})}\BibitemShut
  {NoStop}%
\bibitem [{\citenamefont {Capel}\ \emph {et~al.}(2010)\citenamefont {Capel},
  \citenamefont {Hussein},\ and\ \citenamefont {Baye}}]{CHB10}%
  \BibitemOpen
  \bibfield  {author} {\bibinfo {author} {\bibfnamefont {P.}~\bibnamefont
  {Capel}}, \bibinfo {author} {\bibfnamefont {M.}~\bibnamefont {Hussein}}, \
  and\ \bibinfo {author} {\bibfnamefont {D.}~\bibnamefont {Baye}},\ }\href
  {\doibase https://doi.org/10.1016/j.physletb.2010.08.072} {\bibfield
  {journal} {\bibinfo  {journal} {Phys. Lett.}\ }\textbf {\bibinfo {volume}
  {B693}},\ \bibinfo {pages} {448} (\bibinfo {year} {2010})}\BibitemShut
  {NoStop}%
\bibitem [{\citenamefont {Johnson}\ \emph {et~al.}(1997)\citenamefont
  {Johnson}, \citenamefont {Al-Khalili},\ and\ \citenamefont
  {Tostevin}}]{JAT97}%
  \BibitemOpen
  \bibfield  {author} {\bibinfo {author} {\bibfnamefont {R.~C.}\ \bibnamefont
  {Johnson}}, \bibinfo {author} {\bibfnamefont {J.~S.}\ \bibnamefont
  {Al-Khalili}}, \ and\ \bibinfo {author} {\bibfnamefont {J.}~\bibnamefont
  {Tostevin}},\ }\href {\doibase 10.1103/PhysRevLett.79.2771} {\bibfield
  {journal} {\bibinfo  {journal} {Phys. Rev. Lett.}\ }\textbf {\bibinfo
  {volume} {79}},\ \bibinfo {pages} {2771} (\bibinfo {year}
  {1997})}\BibitemShut {NoStop}%
\bibitem [{\citenamefont {Johnson}(1999)}]{proc97}%
  \BibitemOpen
  \bibfield  {author} {\bibinfo {author} {\bibfnamefont {R.~C.}\ \bibnamefont
  {Johnson}},\ }in\ \href@noop {} {\emph {\bibinfo {booktitle} {Proc. of the
  Euro. Conf. in Advances in Nucl. Phys. and Related Areas {\rm (July 1997,
  Thessaloniki, Greece)}}}},\ \bibinfo {editor} {edited by\ \bibinfo {editor}
  {\bibfnamefont {D.}~\bibnamefont {Brink}}, \bibinfo {editor} {\bibfnamefont
  {M.}~\bibnamefont {Grypeos}}, \ and\ \bibinfo {editor} {\bibfnamefont
  {S.}~\bibnamefont {Massen}}}\ (\bibinfo  {publisher} {Giahoudi-Giapouli
  Publishing},\ \bibinfo {address} {Thessaloniki},\ \bibinfo {year} {1999})\
  p.\ \bibinfo {pages} {156}\BibitemShut {NoStop}%
\bibitem [{\citenamefont {Capel}\ \emph {et~al.}(2011)\citenamefont {Capel},
  \citenamefont {Johnson},\ and\ \citenamefont {Nunes}}]{CJN11}%
  \BibitemOpen
  \bibfield  {author} {\bibinfo {author} {\bibfnamefont {P.}~\bibnamefont
  {Capel}}, \bibinfo {author} {\bibfnamefont {R.}~\bibnamefont {Johnson}}, \
  and\ \bibinfo {author} {\bibfnamefont {F.}~\bibnamefont {Nunes}},\ }\href
  {\doibase https://doi.org/10.1016/j.physletb.2011.09.105} {\bibfield
  {journal} {\bibinfo  {journal} {Phys. Lett.}\ }\textbf {\bibinfo {volume}
  {B705}},\ \bibinfo {pages} {112} (\bibinfo {year} {2011})}\BibitemShut
  {NoStop}%
\bibitem [{\citenamefont {Capel}\ \emph {et~al.}(2013)\citenamefont {Capel},
  \citenamefont {Johnson},\ and\ \citenamefont {Nunes}}]{CJN13}%
  \BibitemOpen
  \bibfield  {author} {\bibinfo {author} {\bibfnamefont {P.}~\bibnamefont
  {Capel}}, \bibinfo {author} {\bibfnamefont {R.~C.}\ \bibnamefont {Johnson}},
  \ and\ \bibinfo {author} {\bibfnamefont {F.~M.}\ \bibnamefont {Nunes}},\
  }\href {\doibase 10.1103/PhysRevC.88.044602} {\bibfield  {journal} {\bibinfo
  {journal} {Phys. Rev. C}\ }\textbf {\bibinfo {volume} {88}},\ \bibinfo
  {pages} {044602} (\bibinfo {year} {2013})}\BibitemShut {NoStop}%
\bibitem [{\citenamefont {Capel}\ \emph {et~al.}(2020)\citenamefont {Capel},
  \citenamefont {Johnson},\ and\ \citenamefont {Nunes}}]{CJN20}%
  \BibitemOpen
  \bibfield  {author} {\bibinfo {author} {\bibfnamefont {P.}~\bibnamefont
  {Capel}}, \bibinfo {author} {\bibfnamefont {R.}~\bibnamefont {Johnson}}, \
  and\ \bibinfo {author} {\bibfnamefont {F.}~\bibnamefont {Nunes}},\ }\href
  {\doibase https://doi.org/10.1140/epja/s10050-020-00310-w} {\bibfield
  {journal} {\bibinfo  {journal} {Eur. Phys. J. A}\ }\textbf {\bibinfo {volume}
  {56}},\ \bibinfo {pages} {300} (\bibinfo {year} {2020})}\BibitemShut
  {NoStop}%
\bibitem [{\citenamefont {Colomer}\ \emph {et~al.}(2016)\citenamefont
  {Colomer}, \citenamefont {Capel}, \citenamefont {Nunes},\ and\ \citenamefont
  {Johnson}}]{CCN16}%
  \BibitemOpen
  \bibfield  {author} {\bibinfo {author} {\bibfnamefont {F.}~\bibnamefont
  {Colomer}}, \bibinfo {author} {\bibfnamefont {P.}~\bibnamefont {Capel}},
  \bibinfo {author} {\bibfnamefont {F.~M.}\ \bibnamefont {Nunes}}, \ and\
  \bibinfo {author} {\bibfnamefont {R.~C.}\ \bibnamefont {Johnson}},\ }\href
  {\doibase 10.1103/PhysRevC.93.054621} {\bibfield  {journal} {\bibinfo
  {journal} {Phys. Rev. C}\ }\textbf {\bibinfo {volume} {93}},\ \bibinfo
  {pages} {054621} (\bibinfo {year} {2016})}\BibitemShut {NoStop}%
\bibitem [{\citenamefont {Yun}\ \emph {et~al.}(2019)\citenamefont {Yun},
  \citenamefont {Colomer}, \citenamefont {Pang},\ and\ \citenamefont
  {Capel}}]{YCP19}%
  \BibitemOpen
  \bibfield  {author} {\bibinfo {author} {\bibfnamefont {X.~Y.}\ \bibnamefont
  {Yun}}, \bibinfo {author} {\bibfnamefont {F.}~\bibnamefont {Colomer}},
  \bibinfo {author} {\bibfnamefont {D.~Y.}\ \bibnamefont {Pang}}, \ and\
  \bibinfo {author} {\bibfnamefont {P.}~\bibnamefont {Capel}},\ }\href
  {\doibase 10.1088/1361-6471/ab355e} {\bibfield  {journal} {\bibinfo
  {journal} {J. Phys. G}\ }\textbf {\bibinfo {volume} {46}},\ \bibinfo {pages}
  {105111} (\bibinfo {year} {2019})}\BibitemShut {NoStop}%
\bibitem [{\citenamefont {Duan}\ \emph {et~al.}(2022)\citenamefont {Duan},
  \citenamefont {Yang}, \citenamefont {Lei}, \citenamefont {Wang},
  \citenamefont {Sun}, \citenamefont {Pang}, \citenamefont {Wang},
  \citenamefont {Liu}, \citenamefont {Xu}, \citenamefont {Ma}, \citenamefont
  {Ma}, \citenamefont {Bai}, \citenamefont {Hu}, \citenamefont {Gao},
  \citenamefont {Xu}, \citenamefont {Lin}, \citenamefont {Jia}, \citenamefont
  {Ma}, \citenamefont {Sun}, \citenamefont {Wang}, \citenamefont {Yang},
  \citenamefont {Jin}, \citenamefont {Ren}, \citenamefont {Zhang},
  \citenamefont {Zhou}, \citenamefont {Hu},\ and\ \citenamefont {Xu}}]{Duan22}%
  \BibitemOpen
  \bibfield  {author} {\bibinfo {author} {\bibfnamefont {F.~F.}\ \bibnamefont
  {Duan}}, \bibinfo {author} {\bibfnamefont {Y.~Y.}\ \bibnamefont {Yang}},
  \bibinfo {author} {\bibfnamefont {J.}~\bibnamefont {Lei}}, \bibinfo {author}
  {\bibfnamefont {K.}~\bibnamefont {Wang}}, \bibinfo {author} {\bibfnamefont
  {Z.~Y.}\ \bibnamefont {Sun}}, \bibinfo {author} {\bibfnamefont {D.~Y.}\
  \bibnamefont {Pang}}, \bibinfo {author} {\bibfnamefont {J.~S.}\ \bibnamefont
  {Wang}}, \bibinfo {author} {\bibfnamefont {X.}~\bibnamefont {Liu}}, \bibinfo
  {author} {\bibfnamefont {S.~W.}\ \bibnamefont {Xu}}, \bibinfo {author}
  {\bibfnamefont {J.~B.}\ \bibnamefont {Ma}}, \bibinfo {author} {\bibfnamefont
  {P.}~\bibnamefont {Ma}}, \bibinfo {author} {\bibfnamefont {Z.}~\bibnamefont
  {Bai}}, \bibinfo {author} {\bibfnamefont {Q.}~\bibnamefont {Hu}}, \bibinfo
  {author} {\bibfnamefont {Z.~H.}\ \bibnamefont {Gao}}, \bibinfo {author}
  {\bibfnamefont {X.~X.}\ \bibnamefont {Xu}}, \bibinfo {author} {\bibfnamefont
  {C.~J.}\ \bibnamefont {Lin}}, \bibinfo {author} {\bibfnamefont {H.~M.}\
  \bibnamefont {Jia}}, \bibinfo {author} {\bibfnamefont {N.~R.}\ \bibnamefont
  {Ma}}, \bibinfo {author} {\bibfnamefont {L.~J.}\ \bibnamefont {Sun}},
  \bibinfo {author} {\bibfnamefont {D.~X.}\ \bibnamefont {Wang}}, \bibinfo
  {author} {\bibfnamefont {G.}~\bibnamefont {Yang}}, \bibinfo {author}
  {\bibfnamefont {S.~Y.}\ \bibnamefont {Jin}}, \bibinfo {author} {\bibfnamefont
  {Z.~Z.}\ \bibnamefont {Ren}}, \bibinfo {author} {\bibfnamefont {Y.~H.}\
  \bibnamefont {Zhang}}, \bibinfo {author} {\bibfnamefont {X.~H.}\ \bibnamefont
  {Zhou}}, \bibinfo {author} {\bibfnamefont {Z.~G.}\ \bibnamefont {Hu}}, \ and\
  \bibinfo {author} {\bibfnamefont {H.~S.}\ \bibnamefont {Xu}} (\bibinfo
  {collaboration} {RIBLL Collaboration}),\ }\href {\doibase
  10.1103/PhysRevC.105.034602} {\bibfield  {journal} {\bibinfo  {journal}
  {Phys. Rev. C}\ }\textbf {\bibinfo {volume} {105}},\ \bibinfo {pages}
  {034602} (\bibinfo {year} {2022})}\BibitemShut {NoStop}%
\bibitem [{\citenamefont {Tribble}\ \emph {et~al.}(1989)\citenamefont
  {Tribble}, \citenamefont {Burch},\ and\ \citenamefont
  {Gagliardi}}]{Tribble1989}%
  \BibitemOpen
  \bibfield  {author} {\bibinfo {author} {\bibfnamefont {R.~E.}\ \bibnamefont
  {Tribble}}, \bibinfo {author} {\bibfnamefont {R.~H.}\ \bibnamefont {Burch}},
  \ and\ \bibinfo {author} {\bibfnamefont {C.~A.}\ \bibnamefont {Gagliardi}},\
  }\href {\doibase https://doi.org/10.1016/0168-9002(89)90215-5} {\bibfield
  {journal} {\bibinfo  {journal} {Nucl. Instr. Meth. A}\ }\textbf {\bibinfo
  {volume} {285}},\ \bibinfo {pages} {441} (\bibinfo {year}
  {1989})}\BibitemShut {NoStop}%
\bibitem [{\citenamefont {Ota}\ \emph {et~al.}(2024)\citenamefont {Ota},
  \citenamefont {Christian}, \citenamefont {Reed}, \citenamefont {Catford},
  \citenamefont {Dede}, \citenamefont {Doherty}, \citenamefont {Lotay},
  \citenamefont {Roosa}, \citenamefont {Saastamoinen},\ and\ \citenamefont
  {Scriven}}]{Ota2023}%
  \BibitemOpen
  \bibfield  {author} {\bibinfo {author} {\bibfnamefont {S.}~\bibnamefont
  {Ota}}, \bibinfo {author} {\bibfnamefont {G.}~\bibnamefont {Christian}},
  \bibinfo {author} {\bibfnamefont {B.~J.}\ \bibnamefont {Reed}}, \bibinfo
  {author} {\bibfnamefont {W.~N.}\ \bibnamefont {Catford}}, \bibinfo {author}
  {\bibfnamefont {S.}~\bibnamefont {Dede}}, \bibinfo {author} {\bibfnamefont
  {D.~T.}\ \bibnamefont {Doherty}}, \bibinfo {author} {\bibfnamefont
  {G.}~\bibnamefont {Lotay}}, \bibinfo {author} {\bibfnamefont
  {M.}~\bibnamefont {Roosa}}, \bibinfo {author} {\bibfnamefont
  {A.}~\bibnamefont {Saastamoinen}}, \ and\ \bibinfo {author} {\bibfnamefont
  {D.~P.}\ \bibnamefont {Scriven}},\ }\href {\doibase
  https://doi.org/10.1016/j.nima.2023.168946} {\bibfield  {journal} {\bibinfo
  {journal} {Nucl. Instr. Meth. A}\ }\textbf {\bibinfo {volume} {1059}},\
  \bibinfo {pages} {168946} (\bibinfo {year} {2024})}\BibitemShut {NoStop}%
\bibitem [{\citenamefont {Baye}\ \emph {et~al.}(2005)\citenamefont {Baye},
  \citenamefont {Capel},\ and\ \citenamefont {Goldstein}}]{BCG05}%
  \BibitemOpen
  \bibfield  {author} {\bibinfo {author} {\bibfnamefont {D.}~\bibnamefont
  {Baye}}, \bibinfo {author} {\bibfnamefont {P.}~\bibnamefont {Capel}}, \ and\
  \bibinfo {author} {\bibfnamefont {G.}~\bibnamefont {Goldstein}},\ }\href
  {\doibase 10.1103/PhysRevLett.95.082502} {\bibfield  {journal} {\bibinfo
  {journal} {Phys. Rev. Lett.}\ }\textbf {\bibinfo {volume} {95}},\ \bibinfo
  {pages} {082502} (\bibinfo {year} {2005})}\BibitemShut {NoStop}%
\bibitem [{\citenamefont {Goldstein}\ \emph {et~al.}(2006)\citenamefont
  {Goldstein}, \citenamefont {Baye},\ and\ \citenamefont {Capel}}]{GBC06}%
  \BibitemOpen
  \bibfield  {author} {\bibinfo {author} {\bibfnamefont {G.}~\bibnamefont
  {Goldstein}}, \bibinfo {author} {\bibfnamefont {D.}~\bibnamefont {Baye}}, \
  and\ \bibinfo {author} {\bibfnamefont {P.}~\bibnamefont {Capel}},\ }\href
  {\doibase 10.1103/PhysRevC.73.024602} {\bibfield  {journal} {\bibinfo
  {journal} {Phys. Rev. C}\ }\textbf {\bibinfo {volume} {73}},\ \bibinfo
  {pages} {024602} (\bibinfo {year} {2006})}\BibitemShut {NoStop}%
\bibitem [{\citenamefont {Fukui}\ \emph {et~al.}(2014)\citenamefont {Fukui},
  \citenamefont {Ogata},\ and\ \citenamefont {Capel}}]{FOC14}%
  \BibitemOpen
  \bibfield  {author} {\bibinfo {author} {\bibfnamefont {T.}~\bibnamefont
  {Fukui}}, \bibinfo {author} {\bibfnamefont {K.}~\bibnamefont {Ogata}}, \ and\
  \bibinfo {author} {\bibfnamefont {P.}~\bibnamefont {Capel}},\ }\href
  {\doibase 10.1103/PhysRevC.90.034617} {\bibfield  {journal} {\bibinfo
  {journal} {Phys. Rev. C}\ }\textbf {\bibinfo {volume} {90}},\ \bibinfo
  {pages} {034617} (\bibinfo {year} {2014})}\BibitemShut {NoStop}%
\bibitem [{\citenamefont {Calci}\ \emph {et~al.}(2016)\citenamefont {Calci},
  \citenamefont {Navr\'atil}, \citenamefont {Roth}, \citenamefont
  {Dohet-Eraly}, \citenamefont {Quaglioni},\ and\ \citenamefont
  {Hupin}}]{Cal16}%
  \BibitemOpen
  \bibfield  {author} {\bibinfo {author} {\bibfnamefont {A.}~\bibnamefont
  {Calci}}, \bibinfo {author} {\bibfnamefont {P.}~\bibnamefont {Navr\'atil}},
  \bibinfo {author} {\bibfnamefont {R.}~\bibnamefont {Roth}}, \bibinfo {author}
  {\bibfnamefont {J.}~\bibnamefont {Dohet-Eraly}}, \bibinfo {author}
  {\bibfnamefont {S.}~\bibnamefont {Quaglioni}}, \ and\ \bibinfo {author}
  {\bibfnamefont {G.}~\bibnamefont {Hupin}},\ }\href {\doibase
  10.1103/PhysRevLett.117.242501} {\bibfield  {journal} {\bibinfo  {journal}
  {Phys. Rev. Lett.}\ }\textbf {\bibinfo {volume} {117}},\ \bibinfo {pages}
  {242501} (\bibinfo {year} {2016})}\BibitemShut {NoStop}%
\bibitem [{\citenamefont {Capel}\ \emph {et~al.}(2018)\citenamefont {Capel},
  \citenamefont {Phillips},\ and\ \citenamefont {Hammer}}]{CPH18}%
  \BibitemOpen
  \bibfield  {author} {\bibinfo {author} {\bibfnamefont {P.}~\bibnamefont
  {Capel}}, \bibinfo {author} {\bibfnamefont {D.~R.}\ \bibnamefont {Phillips}},
  \ and\ \bibinfo {author} {\bibfnamefont {H.-W.}\ \bibnamefont {Hammer}},\
  }\href {\doibase 10.1103/PhysRevC.98.034610} {\bibfield  {journal} {\bibinfo
  {journal} {Phys. Rev. C}\ }\textbf {\bibinfo {volume} {98}},\ \bibinfo
  {pages} {034610} (\bibinfo {year} {2018})}\BibitemShut {NoStop}%
\bibitem [{\citenamefont {Moschini}\ and\ \citenamefont {Capel}(2019)}]{MC19}%
  \BibitemOpen
  \bibfield  {author} {\bibinfo {author} {\bibfnamefont {L.}~\bibnamefont
  {Moschini}}\ and\ \bibinfo {author} {\bibfnamefont {P.}~\bibnamefont
  {Capel}},\ }\href {\doibase https://doi.org/10.1016/j.physletb.2019.01.041}
  {\bibfield  {journal} {\bibinfo  {journal} {Phys. Lett.}\ }\textbf {\bibinfo
  {volume} {B790}},\ \bibinfo {pages} {367} (\bibinfo {year}
  {2019})}\BibitemShut {NoStop}%
\bibitem [{\citenamefont {Yang}\ and\ \citenamefont {Capel}(2018)}]{YC18}%
  \BibitemOpen
  \bibfield  {author} {\bibinfo {author} {\bibfnamefont {J.}~\bibnamefont
  {Yang}}\ and\ \bibinfo {author} {\bibfnamefont {P.}~\bibnamefont {Capel}},\
  }\href {\doibase 10.1103/PhysRevC.98.054602} {\bibfield  {journal} {\bibinfo
  {journal} {Phys. Rev. C}\ }\textbf {\bibinfo {volume} {98}},\ \bibinfo
  {pages} {054602} (\bibinfo {year} {2018})}\BibitemShut {NoStop}%
\bibitem [{\citenamefont {Hebborn}\ and\ \citenamefont {Capel}(2021)}]{HC21}%
  \BibitemOpen
  \bibfield  {author} {\bibinfo {author} {\bibfnamefont {C.}~\bibnamefont
  {Hebborn}}\ and\ \bibinfo {author} {\bibfnamefont {P.}~\bibnamefont
  {Capel}},\ }\href {\doibase 10.1103/PhysRevC.104.024616} {\bibfield
  {journal} {\bibinfo  {journal} {Phys. Rev. C}\ }\textbf {\bibinfo {volume}
  {104}},\ \bibinfo {pages} {024616} (\bibinfo {year} {2021})}\BibitemShut
  {NoStop}%
\bibitem [{\citenamefont {Bertulani}\ \emph {et~al.}(2002)\citenamefont
  {Bertulani}, \citenamefont {Hammer},\ and\ \citenamefont {{van
  Kolck}}}]{BHvK02}%
  \BibitemOpen
  \bibfield  {author} {\bibinfo {author} {\bibfnamefont {C.}~\bibnamefont
  {Bertulani}}, \bibinfo {author} {\bibfnamefont {H.-W.}\ \bibnamefont
  {Hammer}}, \ and\ \bibinfo {author} {\bibfnamefont {U.}~\bibnamefont {{van
  Kolck}}},\ }\href {\doibase https://doi.org/10.1016/S0375-9474(02)01270-8}
  {\bibfield  {journal} {\bibinfo  {journal} {Nucl. Phys.}\ }\textbf {\bibinfo
  {volume} {A712}},\ \bibinfo {pages} {37} (\bibinfo {year}
  {2002})}\BibitemShut {NoStop}%
\bibitem [{\citenamefont {Hammer}\ \emph {et~al.}(2017)\citenamefont {Hammer},
  \citenamefont {Ji},\ and\ \citenamefont {Phillips}}]{HJP17}%
  \BibitemOpen
  \bibfield  {author} {\bibinfo {author} {\bibfnamefont {H.-W.}\ \bibnamefont
  {Hammer}}, \bibinfo {author} {\bibfnamefont {C.}~\bibnamefont {Ji}}, \ and\
  \bibinfo {author} {\bibfnamefont {D.~R.}\ \bibnamefont {Phillips}},\ }\href
  {\doibase 10.1088/1361-6471/aa83db} {\bibfield  {journal} {\bibinfo
  {journal} {J. Phys. G}\ }\textbf {\bibinfo {volume} {44}},\ \bibinfo {pages}
  {103002} (\bibinfo {year} {2017})}\BibitemShut {NoStop}%
\bibitem [{\citenamefont {Al-Khalili}\ \emph {et~al.}(1997)\citenamefont
  {Al-Khalili}, \citenamefont {Tostevin},\ and\ \citenamefont
  {Brooke}}]{ATB97}%
  \BibitemOpen
  \bibfield  {author} {\bibinfo {author} {\bibfnamefont {J.~S.}\ \bibnamefont
  {Al-Khalili}}, \bibinfo {author} {\bibfnamefont {J.~A.}\ \bibnamefont
  {Tostevin}}, \ and\ \bibinfo {author} {\bibfnamefont {J.~M.}\ \bibnamefont
  {Brooke}},\ }\href {\doibase 10.1103/PhysRevC.55.R1018} {\bibfield  {journal}
  {\bibinfo  {journal} {Phys. Rev. C}\ }\textbf {\bibinfo {volume} {55}},\
  \bibinfo {pages} {R1018} (\bibinfo {year} {1997})}\BibitemShut {NoStop}%
\bibitem [{\citenamefont {Durant}\ \emph {et~al.}(2018)\citenamefont {Durant},
  \citenamefont {Capel}, \citenamefont {Huth}, \citenamefont {Balantekin},\
  and\ \citenamefont {Schwenk}}]{Dur18}%
  \BibitemOpen
  \bibfield  {author} {\bibinfo {author} {\bibfnamefont {V.}~\bibnamefont
  {Durant}}, \bibinfo {author} {\bibfnamefont {P.}~\bibnamefont {Capel}},
  \bibinfo {author} {\bibfnamefont {L.}~\bibnamefont {Huth}}, \bibinfo {author}
  {\bibfnamefont {A.}~\bibnamefont {Balantekin}}, \ and\ \bibinfo {author}
  {\bibfnamefont {A.}~\bibnamefont {Schwenk}},\ }\href {\doibase
  https://doi.org/10.1016/j.physletb.2018.05.084} {\bibfield  {journal}
  {\bibinfo  {journal} {Phys. Lett.}\ }\textbf {\bibinfo {volume} {B782}},\
  \bibinfo {pages} {668} (\bibinfo {year} {2018})}\BibitemShut {NoStop}%
\bibitem [{\citenamefont {Durant}\ \emph {et~al.}(2020)\citenamefont {Durant},
  \citenamefont {Capel},\ and\ \citenamefont {Schwenk}}]{DCS20}%
  \BibitemOpen
  \bibfield  {author} {\bibinfo {author} {\bibfnamefont {V.}~\bibnamefont
  {Durant}}, \bibinfo {author} {\bibfnamefont {P.}~\bibnamefont {Capel}}, \
  and\ \bibinfo {author} {\bibfnamefont {A.}~\bibnamefont {Schwenk}},\ }\href
  {\doibase 10.1103/PhysRevC.102.014622} {\bibfield  {journal} {\bibinfo
  {journal} {Phys. Rev. C}\ }\textbf {\bibinfo {volume} {102}},\ \bibinfo
  {pages} {014622} (\bibinfo {year} {2020})}\BibitemShut {NoStop}%
\bibitem [{\citenamefont {Koning}\ and\ \citenamefont
  {Delaroche}(2003)}]{KD03}%
  \BibitemOpen
  \bibfield  {author} {\bibinfo {author} {\bibfnamefont {A.}~\bibnamefont
  {Koning}}\ and\ \bibinfo {author} {\bibfnamefont {J.}~\bibnamefont
  {Delaroche}},\ }\href {\doibase
  https://doi.org/10.1016/S0375-9474(02)01321-0} {\bibfield  {journal}
  {\bibinfo  {journal} {Nucl. Phys.}\ }\textbf {\bibinfo {volume} {A713}},\
  \bibinfo {pages} {231} (\bibinfo {year} {2003})}\BibitemShut {NoStop}%
\bibitem [{\citenamefont {Anne}\ \emph {et~al.}(1994)\citenamefont {Anne},
  \citenamefont {Bimbot}, \citenamefont {Dogny}, \citenamefont {Emling},
  \citenamefont {Guillemaud-Mueller}, \citenamefont {Hansen}, \citenamefont
  {Hornshøj}, \citenamefont {Humbert}, \citenamefont {Jonson}, \citenamefont
  {Keim}, \citenamefont {Lewitowicz}, \citenamefont {Møller}, \citenamefont
  {Mueller}, \citenamefont {Neugart}, \citenamefont {Nilsson}, \citenamefont
  {Nyman}, \citenamefont {Pougheon}, \citenamefont {Riisager}, \citenamefont
  {Saint-Laurent}, \citenamefont {Schrieder}, \citenamefont {Sorlin},
  \citenamefont {Tengblad},\ and\ \citenamefont {Rolander}}]{Anne94}%
  \BibitemOpen
  \bibfield  {author} {\bibinfo {author} {\bibfnamefont {R.}~\bibnamefont
  {Anne}}, \bibinfo {author} {\bibfnamefont {R.}~\bibnamefont {Bimbot}},
  \bibinfo {author} {\bibfnamefont {S.}~\bibnamefont {Dogny}}, \bibinfo
  {author} {\bibfnamefont {H.}~\bibnamefont {Emling}}, \bibinfo {author}
  {\bibfnamefont {D.}~\bibnamefont {Guillemaud-Mueller}}, \bibinfo {author}
  {\bibfnamefont {P.}~\bibnamefont {Hansen}}, \bibinfo {author} {\bibfnamefont
  {P.}~\bibnamefont {Hornshøj}}, \bibinfo {author} {\bibfnamefont
  {F.}~\bibnamefont {Humbert}}, \bibinfo {author} {\bibfnamefont
  {B.}~\bibnamefont {Jonson}}, \bibinfo {author} {\bibfnamefont
  {M.}~\bibnamefont {Keim}}, \bibinfo {author} {\bibfnamefont {M.}~\bibnamefont
  {Lewitowicz}}, \bibinfo {author} {\bibfnamefont {P.}~\bibnamefont {Møller}},
  \bibinfo {author} {\bibfnamefont {A.}~\bibnamefont {Mueller}}, \bibinfo
  {author} {\bibfnamefont {R.}~\bibnamefont {Neugart}}, \bibinfo {author}
  {\bibfnamefont {T.}~\bibnamefont {Nilsson}}, \bibinfo {author} {\bibfnamefont
  {G.}~\bibnamefont {Nyman}}, \bibinfo {author} {\bibfnamefont
  {F.}~\bibnamefont {Pougheon}}, \bibinfo {author} {\bibfnamefont
  {K.}~\bibnamefont {Riisager}}, \bibinfo {author} {\bibfnamefont {M.-G.}\
  \bibnamefont {Saint-Laurent}}, \bibinfo {author} {\bibfnamefont
  {G.}~\bibnamefont {Schrieder}}, \bibinfo {author} {\bibfnamefont
  {O.}~\bibnamefont {Sorlin}}, \bibinfo {author} {\bibfnamefont
  {O.}~\bibnamefont {Tengblad}}, \ and\ \bibinfo {author} {\bibfnamefont
  {K.}~\bibnamefont {Rolander}},\ }\href {\doibase
  https://doi.org/10.1016/0375-9474(94)90142-2} {\bibfield  {journal} {\bibinfo
   {journal} {Nucl. Phys.}\ }\textbf {\bibinfo {volume} {A575}},\ \bibinfo
  {pages} {125} (\bibinfo {year} {1994})}\BibitemShut {NoStop}%
\bibitem [{\citenamefont {Capel}\ \emph {et~al.}(2022)\citenamefont {Capel},
  \citenamefont {Phillips},\ and\ \citenamefont {Hammer}}]{Capel2022}%
  \BibitemOpen
  \bibfield  {author} {\bibinfo {author} {\bibfnamefont {P.}~\bibnamefont
  {Capel}}, \bibinfo {author} {\bibfnamefont {D.}~\bibnamefont {Phillips}}, \
  and\ \bibinfo {author} {\bibfnamefont {H.-W.}\ \bibnamefont {Hammer}},\
  }\href {\doibase https://doi.org/10.1016/j.physletb.2021.136847} {\bibfield
  {journal} {\bibinfo  {journal} {Phys. Lett. B}\ }\textbf {\bibinfo {volume}
  {825}},\ \bibinfo {pages} {136847} (\bibinfo {year} {2022})}\BibitemShut
  {NoStop}%
\bibitem [{\citenamefont {Durant}\ and\ \citenamefont {Capel}(2024)}]{DC24}%
  \BibitemOpen
  \bibfield  {author} {\bibinfo {author} {\bibfnamefont {V.}~\bibnamefont
  {Durant}}\ and\ \bibinfo {author} {\bibfnamefont {P.}~\bibnamefont {Capel}},\
  }\href@noop {} {\enquote {\bibinfo {title} {Analysis of $^{11}${Be} reactions
  using chiral effective field theory {$NN$} interactions},}\ } (\bibinfo
  {year} {2024}),\ \bibinfo {note} {in preparation}\BibitemShut {NoStop}%
\bibitem [{\citenamefont {Moro}\ and\ \citenamefont {Lay}(2012)}]{ML12}%
  \BibitemOpen
  \bibfield  {author} {\bibinfo {author} {\bibfnamefont {A.~M.}\ \bibnamefont
  {Moro}}\ and\ \bibinfo {author} {\bibfnamefont {J.~A.}\ \bibnamefont {Lay}},\
  }\href {\doibase 10.1103/PhysRevLett.109.232502} {\bibfield  {journal}
  {\bibinfo  {journal} {Phys. Rev. Lett.}\ }\textbf {\bibinfo {volume} {109}},\
  \bibinfo {pages} {232502} (\bibinfo {year} {2012})}\BibitemShut {NoStop}%
\bibitem [{PAC()}]{PAC3}%
  \BibitemOpen
  \href@noop {} {}\bibinfo {howpublished}
  {\url{https://userportal.frib.msu.edu/Pac/Experiments/PublicList}},\ \bibinfo
  {note} {accessed: 2025-03-20}\BibitemShut {NoStop}%
\end{thebibliography}%

\end{document}